\documentclass[twocolumn]{aastex62}

\graphicspath{{./}{figures/}}

\received{2020 May 29}
\revised{2020 June 26}
\accepted{2020 June 27}
\submitjournal{ApJ}

\shorttitle{The Vega Planetary System Resolved by ALMA}
\shortauthors{}

\begin{document}

\title{Dust Populations in the Iconic Vega Planetary System Resolved by ALMA}

\author[0000-0003-4705-3188]{Luca Matr\`a}
\affiliation{Center for Astrophysics $\vert$ Harvard \& Smithsonian, 60 Garden St., Cambridge, MA 02138, USA}
\affiliation{School of Physics, National University of Ireland Galway, University Road, Galway, Ireland}

\author{William R. F. Dent}
\affiliation{ALMA JAO, Alonso de C\'ordova 3107, Casilla 763 0355, Santiago, Chile}
\affiliation{European Southern Observatory, Alonso de C\'ordova 3107, Casilla 763 0355, Santiago, Chile}

\author{David J. Wilner}
\affiliation{Center for Astrophysics $\vert$ Harvard \& Smithsonian, 60 Garden St., Cambridge, MA 02138, USA}

\author{Sebasti\'an Marino}
\affiliation{Max Planck Institute for Astronomy, K\"onigstuhl 17, 69117 Heidelberg, Germany}

\author{Mark C. Wyatt}
\affil{Institute of Astronomy, University of Cambridge, Madingley Road, Cambridge CB3 0HA, UK}

\author{Jonathan P. Marshall}
\affiliation{Academia Sinica Institute of Astronomy and Astrophysics, AS/NTU Astronomy-Mathematics Building, No.1, Sect. 4, Roosevelt Rd, Taipei 10617, Taiwan}

\author{Kate Y. L. Su}
\affiliation{Steward Observatory, University of Arizona, 933 North Cherry Avenue, Tucson, AZ 85721, USA}

\author{Miguel Chavez}
\affiliation{Instituto Nacional de Astrof\'isica, \'Optica y Electr\'onica, Luis Enrique Erro 1, Santa Maria Tonantzintla, Puebla, Mexico}

\author{Antonio Hales}
\affiliation{ALMA JAO, Alonso de C\'ordova 3107, Casilla 763 0355, Santiago, Chile}
\affiliation{National Radio Astronomy Observatory, 520 Edgemont Road, Charlottesville, VA 22903-2475, USA}

\author{A. Meredith Hughes}
\affiliation{Department of Astronomy, Van Vleck Observatory, Wesleyan University, 96 Foss Hill Drive, Middletown, CT 06459, USA}

\author{Jane S. Greaves}
\affiliation{School of Physics and Astronomy, Cardiff University, 4 The Parade, Cardiff CF24 3AA, UK}

\author{Stuartt A. Corder}
\affiliation{ALMA JAO, Alonso de C\'ordova 3107, Casilla 763 0355, Santiago, Chile}
\affiliation{National Radio Astronomy Observatory, 520 Edgemont Road, Charlottesville, VA 22903-2475, USA}







\begin{abstract}

The Vega planetary system hosts the archetype of extrasolar Kuiper belts, and is rich in dust from the sub-au region out to 100's of au, suggesting intense dynamical activity. We present ALMA mm observations that detect and resolve the outer dust belt from the star for the first time. The interferometric visibilities show that the belt can be fit by a Gaussian model or by power-law models with a steep inner edge (at 60-80 au). The belt is very broad, extending out to at least 150-200 au. We strongly detect the star and set a stringent upper limit to warm dust emission previously detected in the infrared. We discuss three scenarios that could explain the architecture of Vega's planetary system, including the new {ALMA} constraints: no outer planets, a chain of low-mass planets, and a single giant planet. The planet-less scenario is only feasible if the outer belt was born with the observed sharp inner edge.  If instead the inner edge is currently being truncated by a planet, then the planet must be $\gtrsim$6 M$_{\oplus}$ and at $\lesssim71$ au to have cleared its chaotic zone within the system age. In the planet chain scenario, outward planet migration and inward scattering of planetesimals could produce the hot and warm dust observed in the inner regions of the system. In the single giant planet scenario, an asteroid belt could be responsible for the warm dust, and mean motion resonances with the planet could put asteroids on star-grazing orbits, producing the hot dust.

\end{abstract}



\section{Introduction} \label{sec:intro}

Debris disks are the dusty signatures of the collisional destruction of planetesimals, exocomets and pebbles around main sequence (and other) stars \citep[see e.g.][and references therein]{Hughes2018}.
The nearby A-type star Vega was one of the first such systems, identified by IRAS through excess emission above the photosphere at 25-100 $\mu $m \citep{Aumann1984}. It is now recognised that at least 24\% of A-type stars harbour such debris \citep{thureau2014}. The spectral energy distribution (SED) of the excess in most cases indicates a temperature of a few tens of K, from dust grains at a few tens of au from the star. As this is also a region where planets may exist, resolved dust images are of considerable interest to help understand the structure and evolution of the co-located planetary systems.

For the closest debris disks, the physical scales imply that they are extended over $\sim$5-20 arcsec, and so the thermal emission can be resolved using single-dish telescopes at mm or far-infrared wavelengths (\citealt{Holland1998,Booth2013,Eiroa2013, Morales2016}). Coronographic images at optical and near-IR wavelengths are also able to resolve some of the more distant systems \cite[e.g.][]{Schneider2014}, and imply the additional presence of small ($\mu$m-sized) scattering grains - likely the end product of the collisional cascade of the larger particles. However, unlike the mm-sized dust, these small grains are strongly affected by radiation pressure and stellar winds, and so the longer wavelength obervations remain the best way to trace the underlying distribution of the parent bodies \citep{Wyatt2006}.

Imaging the mm-wavelength emission from more distant systems typically requires the resolution of interferometers, and these images show that their mm dust can be distributed over a range of different structures. This includes single narrow rings (e.g. HR~4796 and Fomalhaut: \citealt{Kennedy2018, Macgregor2017}), multiple rings with gaps (HD~107146, HD~92945: \citealt{Marino2018b, Marino2019}), broad belts (e.g. HD~95086, HR~8799: \citealt{Su2017, Wilner2018}), extended smooth haloes (HD~32297, HD~61005: \citealt{MacGregor2018}), and vertical substructures ($\beta$ Pictoris: \citealt{Matra2019b}). 

Vega (HD~172167) is an A0V \citep[e.g.][]{GrayGarrison1987} star at a distance of 7.8~pc \citep{vanLeeuwen2007}, with an estimated age of $\sim$400-700 Myr \citep{Yoon2010, Monnier2012}. The star is rapidly rotating, and is viewed almost face-on (i.e. with its poles nearly aligned with the line of sight, with an inclination of $\sim 6^{\circ}$, \citealt{Monnier2012}).

Being the archetypal bright debris disk, Vega has been the subject of intense scrutiny since its detection with IRAS. Follow-up images marginally resolved the emission at mm/sub-mm wavelengths, and suggested a broad, somewhat clumpy structure \citep{Holland1998, Marsh2006}. 
However, further investigations at higher resolution have proved problematic. While initial interferometric results at 1.3 mm also suggested the presence of clumps (at low signal-to-noise), their locations were not consistent with the expectation of Keplerian motion (\citealt{Koerner2001, Wilner2002}). Subsequent non-detections of extended structure at 1.3 and 0.87~mm suggested the dust distribution was actually rather smooth (\citealt{Pietu2011, Hughes2012}). 
Further complicating the picture were the mid- to far-infrared images, which revealed a smooth decrease in intensity with radius, indicative of a 1/r density power law out to $\sim$800~au \citep{Su2005}. With higher resolution (5.6$\arcsec$), evidence of a peak in surface brightness at $\sim$100~au was seen at 70 $\mu$m with {\em Herschel}, after subtracting the relatively bright stellar photosphere \citep{Sibthorpe2010}. 

Completing the picture of Vega's debris system, the SED at shorter wavelengths shows two additional dust components interior to the cold, outer belt. An excess in the mid-IR detected by \textit{Spitzer} indicates the presence of warm dust at $\sim$170 K, with an estimated fractional IR luminosity (compared to the star) of $7\times10^{-6}$. Assuming blackbody emission, this would imply a minimum radius of $\sim$14 au \citep{Su2013}. The emission was unresolved so must be interior to 6$\arcsec$, or $\sim$47 au, and was interpreted as evidence for dust produced by a collisionally evolving asteroid belt. Finally, a near-IR excess with temperature $\gtrsim1000$ K was detected at the $\sim1$\% level using $\sim2$ $\mu$m interferometry \citep{Absil2006,Defrere2011}. Faint detections or non-detections of this component at longer mid-IR wavelengths indicate that the emission arises from dust interior to $\sim$0.5 au, suggesting that the region between the hot and warm dust is relatively dust-poor \citep{Mennesson2011, Ertel2018, Ertel2020}. 

In this paper, motivated to resolve the previous mysteries surrounding Vega's outer belt structure, and to attempt resolved detection of the warm dust component, we use the unprecedented sensitivity of the Atacama Large Millimeter/submillimeter Array (ALMA) to observe the Vega system interferometrically at mm wavelengths. In \S\ref{sec:obs}, we describe the combined ALMA and Atacama Compact Array (ACA) observations, including processing and imaging. In \S\ref{sec:resmod}, we present the results of our imaging analysis, model the outer belt and the star by fitting the interferometric visibilities, and search for evidence of the inner warm belt. Then, we discuss our findings in \S\ref{sec:disc} and conclude with a summary in \S\ref{sec:conc}. 

\section{Observations} \label{sec:obs}
The observations were carried out in ALMA band 6 at a mean frequency of 225.2~GHz (project 2015.1.00182.S). Executions were carried out on both the ACA and the 12m array in the most compact configuration. The total on-source time was 80 minutes over three executions using the 12m array in April 2016, and 400 minutes in 12 executions with the ACA in Aug-Sept 2016. In all observations, J1751+0939 was used for bandpass and flux calibration and J1848+3219 was the phase calibrator. The total bandwidth was 7.5GHz, with three spectral windows set for low resolution TDM mode, and one covering the region of the $^{12}$CO line at 230.538~GHz with a resolution of 1.3 km s$^{-1}$.
The high source declination means that the elevations of the Vega scans were low - in the range 21-28$^{\circ}$. With the ACA and ALMA 12m compact configuration, this meant a considerable amount of antenna shadowing, but this was taken into account by flagging in the data reduction pipeline. Weather conditions were good during most observations, with a mean pwv of $\sim $1~mm (zenith).

In order to improve the image fidelity of the 12m array observations, we self-calibrated the data separately for each of the three executions using the bright, unresolved emission from the central star. We executed 3 rounds of phase-only self-calibration, combining visibilities from all 4 SPWs. We used the stellar point-like emission as a model (outputted by the \textit{tclean} task), and decreased the length of the solution interval ($\sim$370, 132, and 99 seconds) at each phase-only self-cal iteration. Finally, we applied one round of amplitude self-calibration with a $\sim$370 s solution interval. For all observing dates, this procedure successfully led to a $\sim$40-50\% increase in the peak SNR of the star.

Following self-calibration, we shifted the phase center of each execution of the 12m observations to the exact location of the star. We determined the latter from model fitting for the star as a point source in the \textit{u-v} plane using the longest baselines, free of disk emission (\S\ref{sec:starmodel}, and Table \ref{tab:starfit}). We apply these shifts to correct for the presence of non-zero offsets that differ between different 12-m executions, likely due to inaccuracies in the phase referencing of the observations. For ALMA, these are normally less than 15\% of the synthesised beam\footnote{e.g. ALMA Technical Handbook, Chapter 10.5.2}, which is indeed the case for our 12m data (Table \ref{tab:starfit}).

The model fits to the star also indicated a different stellar flux density for each of the 12m executions (Table \ref{tab:starfit}), with the largest difference between any two datasets being $\sim$12\%. This is not significant given the expected 5-10\% flux calibration accuracy in Band 6\footnote{e.g. ALMA Technical Handbook, Chapter 10.2.6}, which is likely underestimated given that our target was observed at low elevations (see above). Therefore, this variability remains fully consistent with instrumental effects alone. To ensure a common flux scale for all the 12m datasets, we adopt the mean of the fluxes as the true stellar flux, and rescale the amplitudes of the 12m datasets to produce a stellar flux of this value.

For the ACA observations, the SNR of the star is insufficient to accurately determine its offset from the phase center and its flux density for each observing date; therefore, we determine the offset by fitting the star to all the executions combined (and with a model including the outer belt, see \S\ref{sec:beltmodel}). Then, as done for the 12m datasets, we shift the phase center and rescale the amplitudes before imaging and further modelling. Note that the phase shifts obtained for the ACA data are consistent with zero within the uncertainties (Table \ref{tab:starfit}).

We imaged the continuum from the combined 12m+ACA phase-shifted and amplitude-rescaled visibility dataset using the \textit{tclean} task within CASA v5.4.0. We employed multi-frequency synthesis with multiscale deconvolution to best recover emission on extended scales from the outer belt, down to point-like emission from the star. In particular, we used scales of 0 (point source), 1, 3, and 9 times the expected synthesized beam size of the image.
We produce two sets of images, one which includes the stellar emission, and one with the stellar emission subtracted from the data in the \textit{u-v} plane, using the best-fit point source model to the stellar emission from \S\ref{sec:starmodel}.

\begin{figure*}
\vspace{-0mm}
 \centering
 \hspace{-15mm}
   \includegraphics*[scale=0.45]{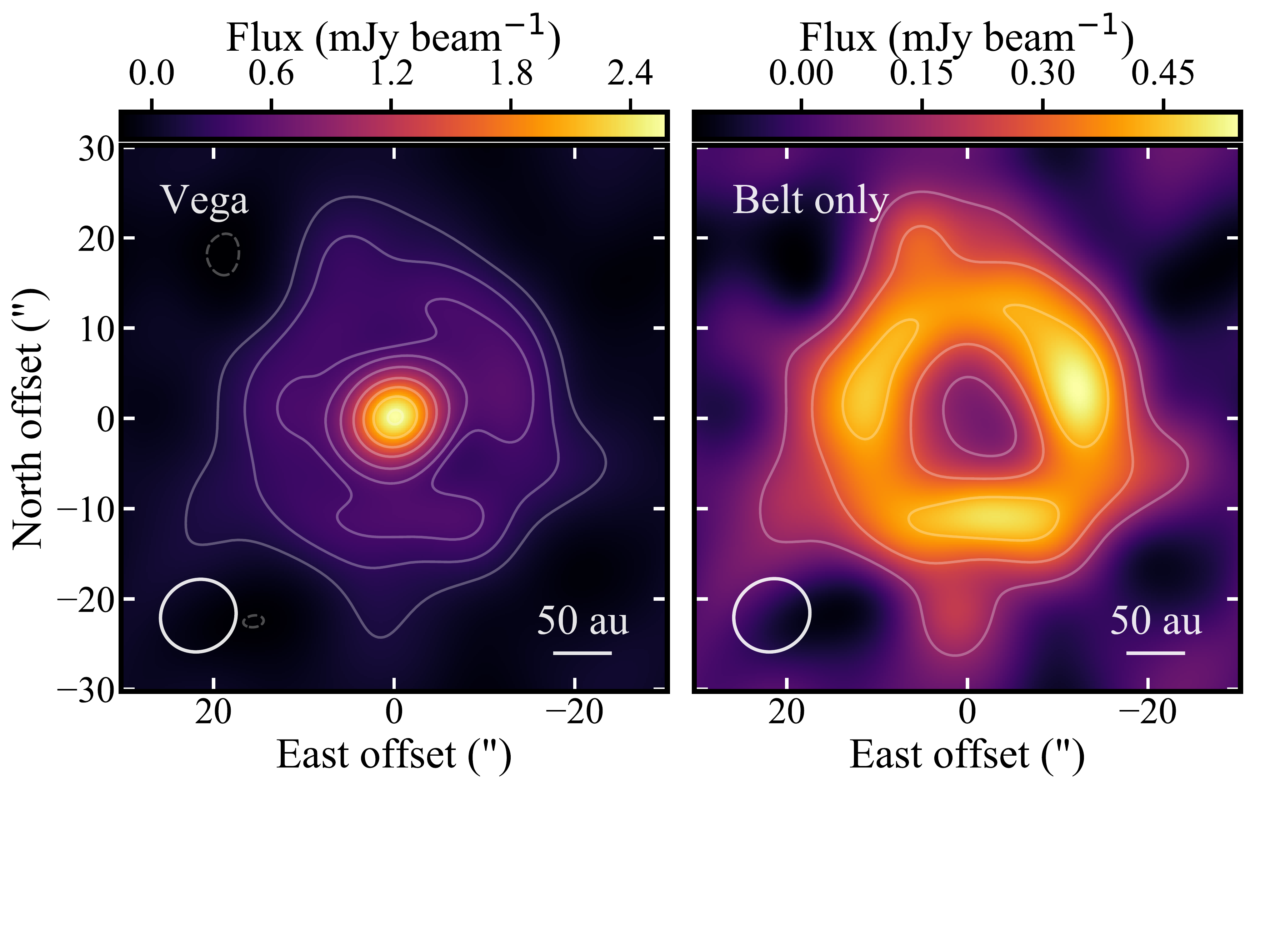}
\vspace{-23mm}
\caption{\textit{Left:} Combined image of the Vega system at 1.34 mm using naturally weighted visibility data from both the ACA and 12m datasets, after applying a 10$\arcsec$ \textit{u-v} taper to enhance the sensitivity to large-scale emission. 
\textit{Right:} Same as left, but with imaging carried out after interferometrically removing the star from the visibilities, as described in \S\ref{sec:starmodel}. Contours are [2,4,6 ..]$\times70\ \mu$Jy beam$^{-1}$, the RMS noise level of the images. No primary beam correction was applied to these images.}
\label{fig:images}
\end{figure*}

Additionally, for each set, we image the data with two different strategies. First, we produce an image of the combined 12m+ACA data with natural weighting, but with a 10$\arcsec$ \textit{u-v} taper applied (Fig. \ref{fig:images}). This maximizes the sensitivity to large scale structure from the outer belt. This image achieves an RMS sensitivity of 70 $\mu$Jy/beam for a synthesized beam of 8.5$\arcsec$ $\times$ 7.9$\arcsec$ with a position angle (PA) of -60.7$^{\circ}$. Then, we image the combined 12m+ACA data with natural visibility weighting (which resolves out the low-surface brightness outer belt) and no taper, to enhance compact emission from the inner region of the system (Fig. \ref{fig:profiles}, top left). This image has an RMS sensitivity of 13 $\mu$Jy/beam for a synthesized beam of 1.8$\arcsec$ $\times$ 0.9$\arcsec$ with a position angle (PA) of -20.8$^{\circ}$.  

For the $^{12}$CO line emission, we extracted the spectral window covering the line frequency from each of the 12m and ACA datasets. We then carried out continuum subtraction in the \textit{u-v} plane using the CASA \textit{uvcontsub} task, avoiding a region $\pm$50 km/s from the line frequency in the rest frame of Vega \citep[accounting for its -20.6$\pm$0.2 km/s heliocentric velocity,][]{Gontcharov2006}. After concatenating all the 12m and ACA visibility datasets, we carried out imaging using natural weighting. The resulting image cube reached a sensitivity of 1 mJy/beam in a 0.64 km/s channel and for a beam size of 1.7$\arcsec$ $\times$ 0.8$\arcsec$, with PA of -20.9$^{\circ}$. No line emission was clearly detected, even after imaging using the same tapering as the continuum observations; therefore, no CLEAN deconvolution was carried out.

\section{Results and Modelling} \label{sec:resmod}
\subsection{Image analysis}
\label{sec:imageanal}

Figure \ref{fig:images} presents the first interferometric detection of Vega's outer belt at mm wavelengths. The 1.34 mm continuum is seen as a ring of emission close to face-on, with a resolved inner hole and a surface brightness peaking at a SNR of $\sim$7 at a distance of 11$\arcsec$-12$\arcsec$ (85-92 au) from the star. The star itself is strongly detected with a peak flux density of 2580 $\mu$Jy (SNR$\sim$37) in the tapered map. 

The map presented is not primary-beam-corrected, and is thus biased toward the inner regions, near the center of the primary beam where the 12m and ACA antennas are most sensitive. To correct for this, we assume that the primary beams resemble an azimuthally symmetric Airy disk (as adopted by CASA), which implies that the sensitivity of the 12m antennas is reduced to 50\% of its maximum at the half-power distance of $\sim$13.3$\arcsec$ ($\sim102$ au), just beyond the peak surface brightness radius of the belt. For the smaller ACA antennas, this 50\% power level is reached at $\sim$23.2$\arcsec$ ($\sim179$ au) from the star.
We note that these CASA-adopted model beams are an adequate description of the real antenna illumination patterns only out to the $\sim$20\% power level\footnote{\url{https://library.nrao.edu/public/memos/naasc/NAASC_117.pdf}}; therefore, we focus on our results within 33.4$\arcsec$ ($\sim257$ au) of the star for the 7m antennas, and within 19.6$\arcsec$ ($\sim151$ au) for the 12m antennas. To calculate the ACA+12m primary beam for joint imaging, we use the A-Projection algorithm as implemented in CASA \footnote{\url{https://casa.nrao.edu/casadocs/casa-5.4.0/synthesis-imaging/mosaicing}}, which accounts for the different primary beams (and their relative weights) in the \textit{u-v} plane by convolving visibilities with the Fourier Transform of the primary beams. 

\begin{figure*}
\vspace{-2mm}
 \hspace{-27mm}
   \includegraphics*[scale=0.62]{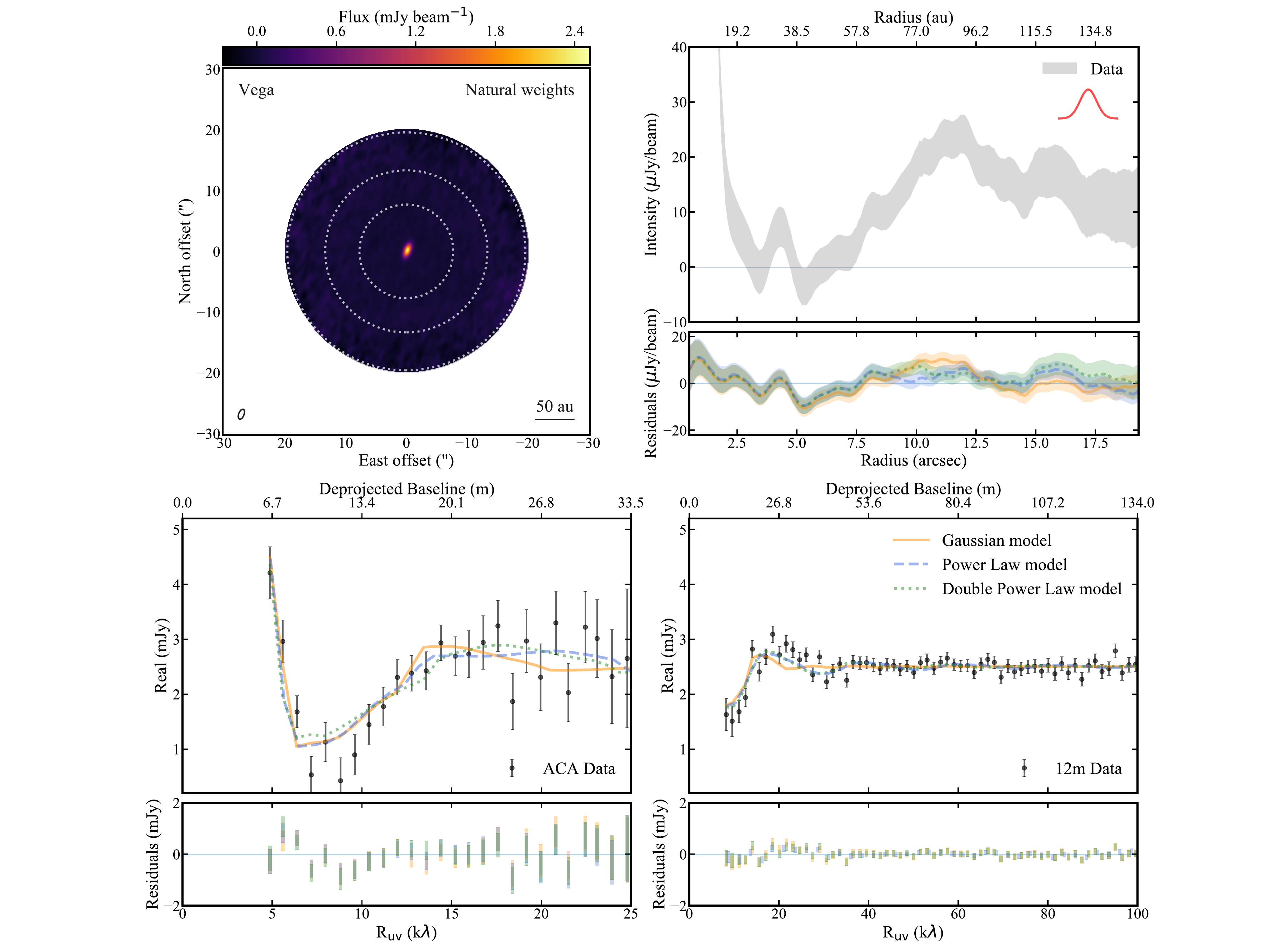}
\vspace{-5mm}
\caption{\textit{Top Left:} Combined image of the Vega system at 1.34 mm using naturally weighted visibility data from both the ACA and 12m datasets. This is the same as Fig. \ref{fig:images} (left), but with no taper applied, and primary-beam-corrected using the combined ACA+12m primary beam. The primary beam's [20$^{\rm th}$,50$^{\rm th}$,80$^{\rm th}$] percentile sensitivity levels are shown as concentric, dotted rings, with sensitivity increasing towards the central star.
\textit{Top Right:} Radial profile of azimuthally averaged emission (upper subpanel) obtained from the Vega naturally weighted untapered image on the top left. The grey area represents the $\pm1\sigma$ range of uncertainty, whereas the red Gaussian represents the beam FWHM of the observations (1.35$\arcsec$, the average between the beam major and minor axis). The lower subpanel shows the radial profile applied to the imaged residuals after subtraction of the best fit models (colors and linestyles). Residual profiles are mostly consistent with zero, indicating a good fit for all of the models, and no significant evidence for inner belt emission (\S\ref{sec:innerbelt}). \textit{Bottom:} Real part of the interferometric visibility data as a function of radius in \textit{u-v} space (black points with $\pm1\sigma$ uncertainties), obtained by azimuthal averaging in concentric \textit{u-v} annuli. These are to be compared to model fits (colors and linestyles matching the right panels). Bottom left is model-data comparison for the ACA visibilities, bottom right is for the 12m visibilities. Note that for overlapping baselines, the ACA and 12m have different real parts due to the different primary beams. The bottom subpanels show residual visibility profiles after subtracting the best-fit outer belt models. The lack of significant residuals indicate that the outer belt -only models are a good fit to the data, and there is no significant evidence for the detection of the warm, inner belt (\S\ref{sec:innerbelt}).
} 
\label{fig:profiles}
\end{figure*}

To study the radial structure in detail, we therefore analyse a primary-beam-corrected radial profile of the continuum emission (Fig. \ref{fig:profiles}, top right) from the higher resolution, naturally weighted 12m+ACA image (Fig. \ref{fig:profiles}, top left). This was constructed by measuring the average surface brightness within concentric, circular annuli (given the near face-on orientation of the belt) at increasing distances from the central star. The uncertainty was measured as the RMS of the naturally weighted map (13 $\mu$Jy/beam), corrected for its increase with radius due to the primary beam correction, and divided by the square root of the number of independent beams along the circumference of each annulus. 

The radial profile of the naturally weighted image confirms that the inner radius of the outer belt is clearly resolved from the star, and that the surface brightness distribution of the outer belt peaks at a radius of $\sim$12$\arcsec$ (92 au) and then decreases extending out to at least 19.6$\arcsec$ (151 au), the 20\% power level of the 12m observations. An important cautionary note is that imaging of very extended, low-level structure filling the primary beams of our observation may be subject to artifacts from CLEAN deconvolution and missing short \textit{u-v} spacings (as shown in \S\ref{sec:beltmodel}), which remove a significant amount of extended emission from the images. Therefore, to robustly derive the mm continuum structure of the Vega planetary system and avoid imaging biases, in the following subsections, we opt to model the visibilities directly.\\

\subsection{Modelling the star}
\label{sec:starmodel}

We first model the star as a point source and fit the long baseline ($>50 k\lambda$) visibilities of each of the 12-m datasets independently. This choice of baselines ensures removal of all detectable emission from the outer belt (see visibility profile in Fig. \ref{fig:imagecombo1}, rightmost panel), which could bias our stellar flux measurement. We confirmed the absence of belt emission by checking images produced using these long baselines only, prior to modelling. We fit the model visibilities to the data through a Markov Chain Monte Carlo (MCMC), implemented using the \textsc{EMCEE} v3 package \citep{Foreman-Mackey2013,Foreman-Mackey2019}. We derive the posterior distribution of each parameter (stellar flux density at 1.34 mm $F_{\nu_{\star}}$, stellar RA and Dec offsets dRA and dDec) using the affine-invariant sampler from \citet{GoodmanWeare2010}, starting from uniform priors on all parameters and using a likelihood function proportional to $e^{-\chi^2/2}$ (with $\chi^2$ being the chi-square function). A fourth parameter we fitted for is a scaling factor for the weights delivered by the ALMA pipeline, $w$, which has been shown to be necessary to ensure that the visibility uncertainties are correct in an absolute sense \citep[e.g.][]{Marino2018b,Matra2019b}.

\begin{deluxetable*}{cccccc}
\tablecaption{Stellar Parameters Derived from Visibility Fitting \label{tab:starfit}}
\tablecolumns{6}
\tablewidth{0pt}
\tablehead{\colhead{Parameter} &
\colhead{Unit} &
\colhead{12m Obs. 0$^{a}$} &
\colhead{12m Obs. 1$^{a}$} &
\colhead{12m Obs. 2$^{a}$} &
\colhead{ACA$^{b}$}}
\startdata
$F_{\nu_{\star}}$
 & $\mu$Jy & $2645^{+36}_{-37}$ & $2489^{+25}_{-23}$ & $2350^{+22}_{-22}$ & $2221^{+83}_{-79}$ \\
dRA*cos(Dec)
 & mas & $15^{+5}_{-4}$ & $50^{+4}_{-4}$ & $57^{+4}_{-4}$ & $71^{+104}_{-103}$ \\
dDec
 & mas & $-41^{+8}_{-9}$ & $2^{+7}_{-7}$ & $-4^{+5}_{-6}$ & $-86^{+127}_{-135}$
\enddata
\tablenotetext{a}{Fitting the star only to 12m-array visibilities beyond $>50$ k$\lambda$}
\tablenotetext{b}{Model including the outer belt, fitted to all ACA visibilities.}
\end{deluxetable*}

We ran the MCMC for 2000 steps using 40 walkers, ensuring convergence.
Table \ref{tab:starfit} shows the resulting best-fit stellar parameters (50$^{\rm th\ +34^{\rm th}}_{\ \ \ -34^{\rm th}}$ percentiles of the marginalized posterior probability distributions). Imaging the visibilities obtained from subtraction of the best-fit model from the data produced noise-like maps, free of significant residual emission at the stellar location.

The best-fit flux densities differ between different datasets, which are at most days apart. However, as mentioned in \S\ref{sec:obs}, this variation is within the flux calibration uncertainty, which means we do not need to invoke a physical variation in the stellar emission, which has been observed around other stars \citep[e.g.][]{White2020}. Considering the flux calibration uncertainty, the observed flux densities are consistent with extrapolation of a Kurucz model from IR wavelengths, which predicts 2.1 mJy at 1.3 mm \citep{Hughes2012}. Note that the hot ($\gtrsim1000$ K) dust component amounts to only $\sim$1.3\% of the stellar flux density at K band \citep{Absil2006}. Given the steep size distribution of this component \citep{Defrere2011}, and the $5-10$\% ALMA calibration uncertainty (1$\sigma$), unresolved detection of the hot dust component is impossible at mm wavelengths.

We adopt a stellar flux density equal to the mean of the best-fit values obtained from the separate 12m datasets (2495 $\mu$Jy), and use it to rescale the amplitudes of each 12m visibility dataset to produce the same stellar emission in all datasets. We also use the derived RA and Dec offsets to align all 12m observations to the same phase center, now corresponding to the photocenter of the stellar emission. The stellar emission cannot be disentangled from that of the outer belt within the ACA dataset (baselines $<29.5 k\lambda$), due to the lack of baselines sufficiently long to \textit{completely} filter out extended emission from the belt. Thus, the star needs to be fit simultaneously to the outer belt before phase alignment and amplitude rescaling of the ACA dataset. 
\\

\subsection{Modelling the outer belt}
\label{sec:beltmodel}

\subsubsection{Modelling framework}
\label{sec:modelframe}

To begin with, we employ a star + outer belt model to the ACA dataset only, to determine the ACA-specific best-fit stellar parameters. For this ACA-only fit, and for the later combined ACA+12m fit, we test three different models describing the radial surface density distribution of mm grains. In the first model, the outer belt is parameterized with an axisymmetric Gaussian surface density distribution of radius $r_c$ and FWHM width $\Delta r$ ($=2\sqrt{2\mathrm{ln}(2)}\sigma_{r}$), and a vertically Gaussian density distribution, following Sect. 4.2 of \citet{Matra2019b}. The mass density distribution of grains reads\footnote{Note that in Eq. 1 of \citet{Matra2019b}, $\Sigma_{\mathrm{dust, }r=r_{\rm c}}$ is erroneously denoted as $\rho_0$ despite having units of surface density.}
\begin{equation}
\rho(r,z)=\Sigma_{\mathrm{dust, }r=r_{\rm c}}\ e^{-\frac{(r-r_{\rm c})^2}{2\sigma_r^2}}\times\frac{e^{-\frac{z^2}{2(hr)^2}}}{\sqrt{2\pi}hr},
\end{equation}
with the same meaning as in previous work (e.g. defining $h=H/r$ where H is the scale height, or standard deviation of the vertical density distribution). We also fit a model where the radial surface density distribution is a power law with slope $\alpha$ and sharp (i.e. unresolved) inner and outer edges $r_{\rm in}$ and $r_{\rm out}$,
\begin{equation}
\rho=\Sigma_{\mathrm{dust, }r=r_{\rm in}}\ \left(\frac{r}{r_{\rm in}}\right)^{\alpha}\frac{e^{-\frac{z^2}{2(hr)^2}}}{\sqrt{2\pi}hr}\ \mathrm{for}\ r_{\rm in}<r<r_{\rm out}.
\end{equation}
Finally, we also attempt a model with no sharp inner and outer edges, where the radial surface density distribution is a combination of two power laws, with slopes $\gamma$ and $\alpha$ respectively interior and exterior to radius $r_{\rm c}$,
\begin{equation}
\rho=\Sigma_{\mathrm{dust, }r=r_{\rm c}}\ \left(\frac{r}{r_{\rm c}}\right)^{\gamma}\frac{e^{-\frac{z^2}{2(hr)^2}}}{\sqrt{2\pi}hr}\ \mathrm{for}\ r<r_{\rm c}
\end{equation}
and
\begin{equation}
\rho=\Sigma_{\mathrm{dust, }r=r_{\rm c}}\ \left(\frac{r}{r_{\rm c}}\right)^{\alpha}\frac{e^{-\frac{z^2}{2(hr)^2}}}{\sqrt{2\pi}hr}\ \mathrm{for}\ r>r_{\rm c}.
\end{equation}

For all models, we fixed the aspect ratio $h$ to 0.03, since this will be largely unconstrained given the face-on viewing geometry of the belt. $\Sigma_{\mathrm{dust, }r=r_{\rm c}}$ is a normalization factor for the surface density of observable dust in the belt, directly linked to the belt flux density under the assumption that the dust opacity is constant throughout the belt. Rather than fitting for $\Sigma_{\mathrm{dust, }r=r_{\rm c}}$, we fit for the belt flux density $F_{\nu, \rm belt}$, which can be done as the emission is optically thin. Finally, we assume the dust to follow a $r^{-0.5}$ radial temperature profile typical of centrally-heated blackbody-like emission. 

For each model realization, we first ray-trace the belt's emission at 1.34 mm using RADMC-3D\footnote{\url{http://www.ita.uni-heidelberg.de/~dullemond/software/radmc-3d/}} to produce a model image. In practice, we create and feed 3D dust density and temperature grid files to RADMC-3D; the 3D dust density is free to change at every model iteration as the parameter space is explored, whereas the temperature grid is kept fixed. Since we fit for the flux density rather than the dust mass in model (by rescaling model images after they are produced) the choice of dust opacity (fed through a dust opacity file) and belt mass (affecting the dust density file) that we provide to RADMC-3D do not matter. This is as long as the dust mass and opacity are chosen to be low enough that the emission calculated by RADMC-3D is optically thin. In our case, we choose a very low mass of $10^{-7}$ M$_{\oplus}$, and a grain opacity of 0.42 cm$^2$ g$^{-1}$ at 1.34 mm, though once again we underline that this choice has no effect on our results.

The ray-traced model image produced by RADMC-3D is multiplied by the primary beam of each of the 12m or ACA observations. Then, we use the GALARIO package \citep{Tazzari2018} to compute its Fourier transform and evaluate it at the \textit{u-v} locations sampled by our observations. Finally, we add a point source representing the star directly at the phase center of the model visibilities, and shift the star+belt model visibilities by RA and Dec offsets ($\Delta$RA, $\Delta$Dec) which we leave as free parameters in the fit. We fit this star+belt model to the observed visibilities with MCMC as done for the star-only fits to the 12m data (\S\ref{sec:starmodel}), using 2000 steps and a number of walkers equal to 10 times the number of free parameters.

\subsubsection{Modelling results}
\label{sec:modelres}

In a first step, we fit the Gaussian model to the ACA dataset alone. This produces best-fit \textit{stellar} parameters listed in Table \ref{tab:starfit} (rightmost column), which we use to 
shift the phase center of the ACA observation to be aligned astrometrically with the 12m observations. Additionally, we use the ACA best-fit stellar flux to rescale the amplitude of the ACA observations to match the amplitude scale of the 12m observations.

\begin{figure*}
\vspace{0mm}
 \hspace{-23mm}
   \includegraphics*[scale=0.62]{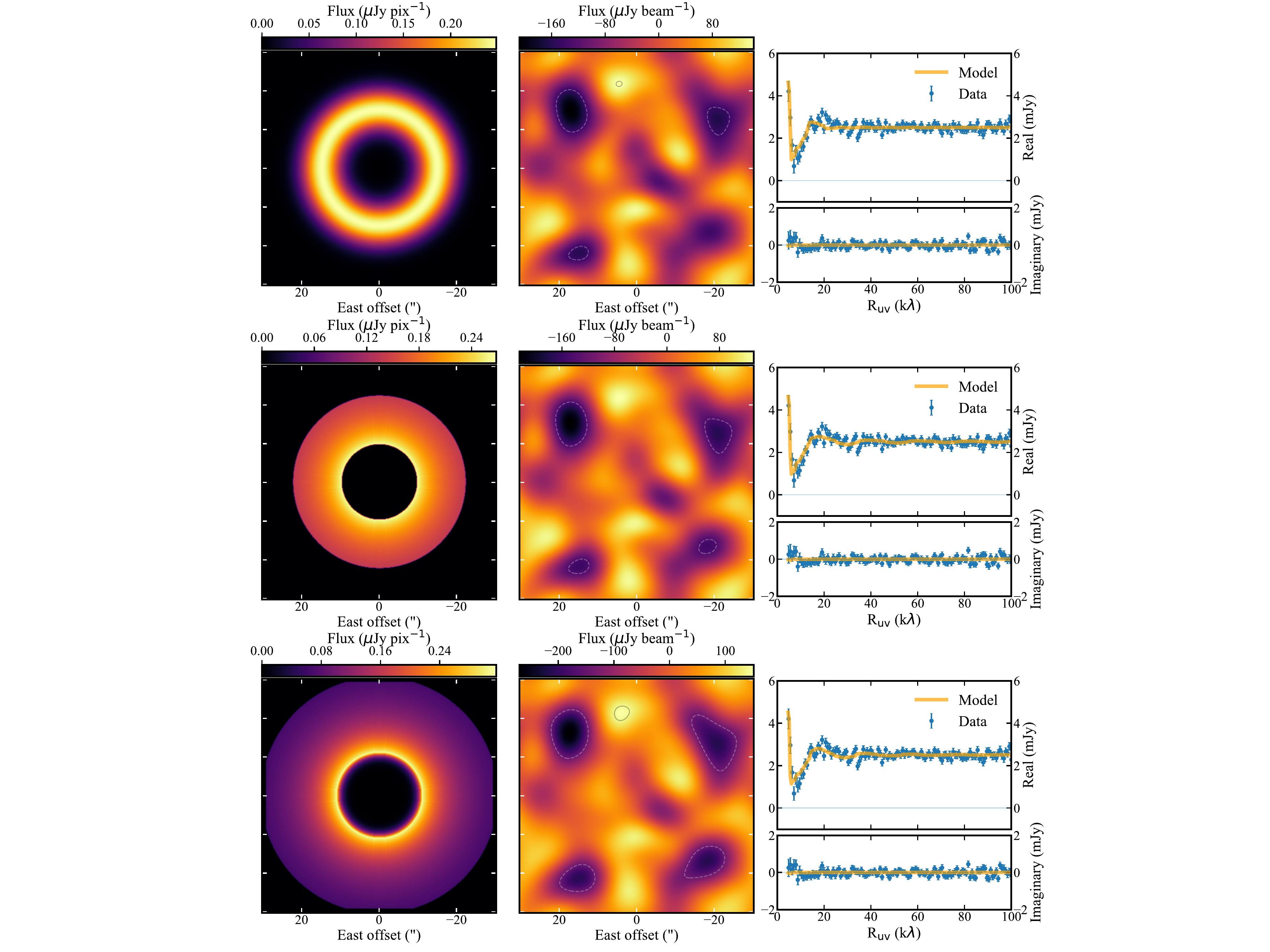}
\vspace{-5mm}
\caption{Joint 12m+ACA modelling results. From left to right column: full resolution best-fit Vega outer belt model images, 10$\arcsec$ tapered images of the residual visibilities (after subtraction of the best-fit model visibilities), and real and imaginary part of the azimuthally averaged complex visibility profiles, for both the data and the best-fit models. The three rows are, from top to bottom, for the Gaussian model, the single power law model, and the double power law model, with best-fit parameters as in Table \ref{tab:beltfit}. Contours in the residual images have the same meaning as in Fig. \ref{fig:images}. 
} 
\label{fig:imagecombo1}
\end{figure*}

In a second step, we fit star+outer belt models to the complete, aligned and amplitude-scaled ACA+12m visibility dataset, which was the one used to produce the images in Fig. \ref{fig:images} as described in \S\ref{sec:obs}. 
The best-fit model parameters (producing the model images shown in Fig. \ref{fig:imagecombo1}, left column) are listed in Table \ref{tab:beltfit}. Residual images, obtained by imaging the residual visibilities with the same imaging parameters as Fig. \ref{fig:images}, are shown in Fig. \ref{fig:imagecombo1} (central column). Finally, the rightmost panels of Fig. \ref{fig:imagecombo1} show the real and imaginary parts of the complex visibilities as a function of deprojected \textit{u-v} distance from the phase center (for both the data and the models). The deprojection has been carried out following the commonly adopted procedure of e.g. \citet{Hughes2007}, assuming an inclination of 0$^{\circ}$ (perfectly face-on), which is consistent with (though not tightly constrained by) the results of our modelling.

\begin{deluxetable*}{ccccccc}
\tablecaption{Outer Belt Model Parameters \label{tab:beltfit}}
\tablecolumns{6}
\tablewidth{0pt}
\tablehead{\colhead{Parameter} &
\colhead{Unit} &
\colhead{Gaussian} &
\colhead{Power-Law} &
\colhead{2 Power-Law}}
\startdata
$F_{\nu_{\rm belt}}$
 & mJy & $8.5^{+1.3}_{-1.1}$ & $8.8^{+1.1}_{-1.0}$ & $12.8^{+1.6}_{-1.5}$ \\
$r_{\rm c}$
 & au ($\arcsec$) & $118^{+4}_{-3}$ ($15.3^{+0.5}_{-0.4}$) & $-$ & $84^{+5}_{-5}$ ($10.9^{+0.6}_{-0.6}$) \\
$\Delta r$
 & au ($\arcsec$) & $67^{+9}_{-7}$ ($8.7^{+1.2}_{-0.9}$) & $-$ & - \\
$r_{\rm in}$
 & au ($\arcsec$) & - & $74^{+3}_{-3}$ ($9.6^{+0.4}_{-0.4}$) & - \\
$r_{\rm out}$
 & au ($\arcsec$) & - & $^{a}171^{+18}_{-11}$ ($22.2^{+2.3}_{-1.4}$) & - \\
$\alpha$
 & - & - & $-0.25^{+0.46}_{-0.51}$ & $-1.0^{+0.3}_{-0.4}$ \\
$\gamma$
 & - & - & - & $10^{+5}_{-3}$ \\
$I$
 & $^{\circ}$ & $^{b}<37$ & $^{b,d}<45$ & $^{b,d}<40$ \\
PA
 & $^{\circ}$ & $^{c}$- & $^{c,d}$- & $^{b,d}-$ \\
Number
 & - & 18 & 19 & 19 \\
$\chi^2$
 & - & 1074973.1 & 1074962.8 & 1074976.3 \\
$^{e}\Delta\chi^2$
 & - & - & -10.3 & +3.2 \\
$^{e}\Delta$AIC
 & - & - & -8.3 & +4.4 \\
$^{e}\Delta$BIC
 & - & - & +2.9 & +16.3\\
\enddata
\tablenotetext{a}{The probability distribuition allows large $r_{\rm out}$ values out to at least 20\% power level of the primary beam, but at a low probability decreasing with radius.}
\tablenotetext{b}{3$\sigma$ upper limit, best-fit value assumed to be 0.0 (face-on)}
\tablenotetext{c}{Probability distribution indicates PA largely unconstrained, though see \textit{d} below.}
\tablenotetext{d}{Significant degeneracy between $I$ and PA. Posterior probability distributions have a peak PA of $\sim45^{\circ}$ preferred for inclinations $\sim25^{\circ}$, approaching our strict upper limit reported. However, PA is increasingly unconstrained for inclinations decreasing down to $0^{\circ}$ (face-on).}
\tablenotetext{e}{With respect to Gaussian model.}
\end{deluxetable*}

By analysing this deprojected real part of the visibility function, we are able to study emission from the outer belt fully independently of the stellar emission. At short baselines or \textit{u-v} distances, the ACA data and the very shortest 12-m baselines reveal the Bessel-function-like visibility function expected from a ring of emission from the outer belt. The belt emission only rises above the star's before the first null of the visibility function, which we find at \textit{u-v} distances of $\sim$5.5 $k\lambda$. This highlights the need for sensitivity on very short baselines, as delivered by the ACA, to recover emission from the outer belt. At all baselines, emission from the star is evident as a constant positive offset of the real component of the visibilities, expected from point-like emission at the phase center of the observations.

This visibility function enables us to attribute the lack of detected emission in previous interferometric observations to spatial filtering and/or a low sensitivity on the shortest baselines. Previous OVRO \citep{Koerner2001}, SMA \citep{Hughes2012} and PdBI \citep{Pietu2011} observations only probed \textit{u-v} distances down to $\sim$11.5 $k\lambda$, thus missing belt emission almost as far as the second null of the visibility function. Previous CARMA observations \citep{Hughes2012} nominally covered \textit{u-v} distances down to $\sim$3.75 $k\lambda$, and therefore could have recovered the emission, but likely did not have sufficient sensitivity to do so at these shortest spacings. Therefore, our detection of the outer belt is consistent with all previously reported interferometric non-detections.

The belt is constrained to have a low inclination, consistent with face-on (in line with the appearance of Fig. \ref{fig:images}), which makes the position angle largely unconstrained given the moderate SNR of the data. The best-fit total belt flux at 1.34 mm varies depending on the model employed since it is based on extrapolating visibility models from the shortest baselines probed by our data to zero \textit{u-v} distance. Including the stellar emission and adding a 15\% flux calibration uncertainty in quadrature, we estimate a model-dependent, total 1.34 mm flux density from the Vega system ranging between $9.9-16.9$ mJy (1$\sigma$ range, including the star) from our ALMA data. This is consistent with recent Large Millimeter Telescope (LMT) and previous sub-mm measurements, which predict a flux density of 13.6~$\pm$~1.6~mJy assuming a spectral slope $\alpha_{\rm mm} = 2.74~\pm~0.33$, calculated from measurements at 450~$\mu$m \citep[229$~\pm~$14~mJy, JCMT/SCUBA-2;][]{Holland2017} and 1100~$\mu$m \citep[21~$\pm$~2~mJy LMT/AzTEC;][]{Marshallinprep}.

No significant emission is seen in the residual 10$\arcsec$-tapered images after subtracting our best-fit axisymmetric models. This corroborates the results of \citet{Hughes2012}, ruling out that compact clumps carry most of the emission within Vega's outer belt. After inspection of residual images obtained with a range of weightings, we find no significant evidence of clumps at all scales from the native resolution of the 12m dataset ($1.7\arcsec\times0.9\arcsec$ with natural weighting, see \S\ref{sec:innerbelt} for details) to the largest scales probed by our shortest baselines ($\sim41\arcsec$).
This is further supported by the lack of significant departures from zero in the imaginary part of the observed visibility function (Fig. \ref{fig:imagecombo1}, right column).

The only residual worthy of note is a compact $\sim$4$\sigma$ peak at $\Delta$RA$\sim0.5\arcsec$, $\Delta$Dec$\sim-10.9\arcsec$ in the naturally weighted (no taper) residual images, south of the star and near the outer belt's inner edge. Assuming it is unresolved, and including flux calibration uncertainty and primary beam correction, its flux density is 81$\pm$20 $\mu$Jy. Taking the galaxy number counts from \citet{Carniani2015}, we expect 3.5 galaxies as bright or brighter than this source within the 20\% power level of the 12-m primary beam. Therefore, we conclude that this source is most likely a background galaxy, although future observations are needed to confirm that this source is not co-moving with Vega.

The lack of significant residuals in the images indicates that all three models (Gaussian, single and double power law) represent the data equally well. Statistical tests such as the Akaike Information Criterion (AIC) and the Bayesian Information Criterion (BIC) confirm that there is no strong evidence in the data that the single and double power law models (which have an extra free parameter) are significantly better at reproducing the data compared to the Gaussian model (Table \ref{tab:beltfit}). This is confirmed visually in Fig. \ref{fig:profiles}, which shows radial profiles in image space (left) and deprojected visibility space (right), including residuals (in the lower panels) consistent with zero across all on-sky and \textit{u-v} radii, for all models.

\subsection{Searching for the inner belt}
\label{sec:innerbelt}

Vega is host to significant mid-infrared excess identified by \citet{Su2013} to be indicative of emission from warm ($\sim$170 K) dust with a fractional luminosity of $\sim7 \times 10^{-6}$. 
If this is black-body emission from a narrow dust belt, it would lie at $\sim$2$\arcsec$ (14 au) radius from the central star. On the other hand, it is well known that small grains dominating the infrared luminosity emit less efficiently than blackbodies, causing true belt radii to be significantly larger \citep[e.g.][]{Booth2013}. Cold belts resolved at mm wavelengths on average show true radii $\sim2.9$ times larger than blackbody radii \citep{Matra2018b}, though with a large scatter, whereas IR observations indicate a stellar luminosity dependence \citep[e.g.][predicting a value of $\sim$1.5 for a 37 L$_{\odot}$ star]{Pawellek2015}. That would imply that Vega's asteroid belt may lie as far as $\sim5.8\arcsec$ ($\sim40$ au) from the star, so we consider its potential location to be anywhere from the $\sim$14 au blackbody radius out to $\sim$40 au. In any case, this warm dust would be spatially resolved in our data.

To avoid potential imaging artifacts due to the strong stellar emission and the extended, partially filtered out cold belt, the best way to look for the warm belt is to analyse the visibility residuals after subtraction of the best-fit star+outer belt model visibilities (Fig. \ref{fig:profiles}, bottom panels).
Assuming a warm belt with a Gaussian radial profile and an unresolved width ($\Delta R\ll1.8\arcsec$ or $\ll14$ au), we expect the real part of the deprojected visibility function (right column in Fig. \ref{fig:profiles}) for the warm belt alone to be well approximated by a Bessel function \citep[e.g.][]{Macgregor2015a}. The first null of the function sets the belt location, via 
\begin{equation}
R_{u-v}=\frac{78.945}{R_{\rm Gaussian}},
\end{equation}
where $R_{u-v}$ is in $k\lambda$ and $R_{\rm Gaussian}$ is in arcseconds. Therefore, in the presence of a narrow, warm belt viewed approximately face-on, and with radius between 2$\arcsec$ and 5.8$\arcsec$, we would expect the residuals of Fig. \ref{fig:profiles}, bottom right to resemble a Bessel function with a null somewhere between $13.6-39.5k\lambda$. No such feature is present at a significant level, although we do note that the residual 12-m visibility function shortward of $\sim$40 k$\lambda$ presents - at a marginal level - hints of substructure at $\gtrsim2\arcsec$ scales not perfectly accounted for by our models.

We can also analyse the residual radial profile obtained after imaging the residual visibilities (Fig. \ref{fig:profiles}, top right, bottom subpanel). Confirming the \textit{u-v} results, no significant emission peak is detected interior to the outer belt, in the $2\arcsec-5.8\arcsec$ region. We can set an upper limit to the 1.34 mm flux density of the warm belt by assuming that its width is unresolved, and that it is azimuthally symmetric. The uncertainty on the azimuthal integral of the intensity in circular annuli with width equal to the beam size $b$ is approximately $\sqrt{N_{\rm b}(R)}\sigma_{\rm b} \sim \sqrt{2\pi R/b}~\sigma_{\rm b}$, where $N_{\rm b}$ is the number of beams along an azimuthal annulus, and where we take $b$ to be the average between the synthesized beam's major and minor axis (1.35$\arcsec$).  
Accounting for primary beam correction, this yields 3$\sigma$ upper limits of $120-223$ $\mu$Jy on the 1.34 mm flux density of a narrow warm belt at radii of $2\arcsec$ to $5.8\arcsec$ (14 to 40 au). Additionally, we can consider warm dust emission to be radially broad rather than constrained to a narrow belt. In that case, we measure a 3$\sigma$ upper limit of $<$324 $\mu$Jy on the total residual 1.34 mm flux density interior to 40 au ($5.8\arcsec$). 

Our 3$\sigma$ upper limits are close to the predicted flux density of 240 $\mu$Jy at 1.34 mm extrapolated from the asteroid belt model of \citet{Su2013} using compact silicates for a belt at 13 au, which fits the \textit{Spitzer} IRS observations. This would imply that the warm emission could have a slightly higher temperature and/or a steeper spectral index than previously assumed, but we conclude that our limit is not sufficiently deep to confirm or rule out the presence of the inner, warm belt. Detailed fitting of the dust spectrum from IR to mm wavelengths, including the inner and outer belt and newly informed by the spatial constraints of the ALMA data will be necessary to reconsider the properties of the system's warm dust excess.

\subsection{CO emission}
\label{sec:gas}
Although undetected in the image cube, we searched for CO emission using the spectro-spatial filtering technique described in \citet{Matra2015,Matra2017b}. In Vega's case, given the face-on configuration, we expect and assume most of the emission to be spectrally unresolved given our spectral resolution of $\sim$1.3 km/s (twice the native channel width of the data). Spatially integrating emission within the outer 4$\sigma$ contour of the continuum observations (Fig. \ref{fig:images}, left), roughly within a circle of about $\sim$130 au radius from the star, yields a spectrum with no significant features, with an RMS of 50 mJy for 0.64 km/s channels. Multiplying by the effective bandwidth of 2.667 times the velocity width of a single channel, we obtain a 3$\sigma$ upper limit on the integrated line flux of 250 mJy km/s. This is a conservative upper limit, as we integrated over a very large surface area within 130 au of the star. Iteratively changing the area to exclude the inner 50 au, or to include only emission within the inner region yielded no significant detection within $\pm$20 km/s of the stellar velocity. 

To turn our integrated line flux upper limit to a CO mass, we employ a non-LTE excitation analysis \citep{Matra2015} including fluorescence due to CO excitation by UV photons \citep{Matra2018a}. In summary, we consider the effect of collisional and radiative excitation by solving the statistical equilibrium equations for the CO molecule, including rotational, vibrational, and electronic levels and transitions. This predicts what fraction of CO molecules will be in the J=2 level of the CO molecule, which in turn allows us to connect our integrated flux upper limit to a CO mass \citep[see e.g. Eq. 2 in][]{Matra2015}.
We assume that emission is optically thin and that collisions with electrons dominate the excitation in a second-generation gas production scenario, where little (if any) H$_2$ would be present. We explore the whole range of electron densities from the radiation- to the collision- dominated regime, and kinetic temperatures between 10 and 250 K, obtaining mass upper limits in the range 3.4-19$\times10^{-7}$ M$_{\oplus}$.

It is important to underline that Vega is a fast rotator, viewed nearly pole-on, but also a luminous early type star with strong UV emission, creating a harsh radiation field for any gas molecules present in the system. For the fluorescent excitation calculation above, and to estimate the CO photodissociation rate, we use the Vega spectrum provided by \citet{Aufdenberg2006} as would be seen along the star's equatorial plane, which differs significantly from that observed from Earth along the polar direction. For example, the estimated luminosity seen by the gas and dust is about 37 L$_{\odot}$, smaller than the bolometric luminosity as observed from Earth ($\sim57$ L$_{\odot}$). We rescale the dust's fractional luminosity to account for this from the observed value of $2\times10^{-5}$ to $3.1\times10^{-5}$. Note that the equatorial Vega spectrum is only provided down to a wavelength of 0.1 $\mu$m, which means we need to extrapolate down to $\sim$0.09 $\mu$m, the shortest wavelength where the stellar UV produces CO photodissociation. We did this by rescaling the flux of a PHOENIX \citep[e.g.][]{Allard2014} spectrum with $T_{\rm eff}=9600$ K, log(g)=4.0 and [M/H]=-0.5, values shown to be a good fit to the observed Vega spectrum \citep{BohlinGilliland2004}, to match the \citet{Aufdenberg2006} spectrum at 0.1 $\mu$m. We use this final spectrum to compute the stellar flux at the radial distance of 118 au (the peak surface density radius of our best-fit Gaussian model) for the fluorescence and photodissociation calculations. This assumes that CO is co-located with the observed dust emission.

We also calculate the CO photodissociation timescale using this spectrum, added to the interstellar radiation field (ISRF), as defined in \citet{Draine1978} with the long-wavelength extension of \citet{vanDishoeck2006}. We find that the stellar UV contribution dominates over the ISRF's in the 900-1100 \AA\ range out to radii of $\sim$350 au, and therefore cannot be neglected as is the case for belts around other stars such as Fomalhaut, despite its spectral type being only a few subtypes later than Vega's. At 118 au, using the cross sections of \citet{Visser2009} as tabulated by \citet{Heays2017}, we calculate a CO photodissociation timescale of 12 years, $\sim10$ times shorter than that expected from UV irradiation by the ISRF alone.

At the estimated age of Vega (400-700 Myr), we would expect any gas, if present, to be produced by second-generation release from exocometary ice. If this release happens at steady state within the collisional cascade that also produces observable dust, and as long as all ice is removed by the time solids are ground down to the smallest grains and removed by radiation pressure, we can use our CO gas mass upper limit to set an upper limit on the ice content of exocometary material \citep[Eq. 2 in][]{Matra2017b}. We assume $r_{\rm c}$ and $\Delta r$ from our best-fit Gaussian model, and a fractional luminosity for the outer belt of $3.1\times10^{-5}$. Additionally, we adopt a mass of 2.135 M$_{\odot}$ \citep{Yoon2010}, and our calculated photodissociation timescale of 12 years. We find that our upper limit does not meaningfully constrain the CO (+CO$_2$, since the latter may also photodissociate and contribute to the observed CO) ice mass fraction of exocomets within the Vega belt. In other words, even if the exocomets' composition was largely dominated by CO and/or CO$_2$, our sensitivity would not have been sufficient to detect gas released by them at steady state within the collisional cascade. For example, if Vega's exocomets had a 10\% CO(+CO$_2$) ice mass fraction, we would have expected a low gas mass of $\sim7.5\times10^{-10}$ M$_{\oplus}$, much below our current upper limits. The low gas mass expected for Vega is mostly driven by its short photodissociation timescale and its low fractional luminosity, the latter implying slow collisional processing and a consequently low CO release rate.

\section{Discussion} \label{sec:disc}

Through the ALMA 12m array and ACA data presented here, we were able to detect Vega's outer belt and measure its spatial properties. We confirm that the belt is very broad with a surface density peak in the 75-120 au region, and that azimuthal clumps do not carry most of the belt's emission. A search for warm dust emission as detected in \textit{Spitzer} and \textit{Herschel} data sets flux density upper limits close to the expectation from extrapolation of the SED model fitted to IR photometry, but do not allow us to conclusively confirm the presence of warm dust in the inner, $\sim$14-40 au region of the system.

With these new constraints, in this Section, we attempt to draw a self-consistent picture on the origin of the architecture of the Vega system, explaining the outer belt structure and accounting for the presence of warm and hot dust emission in the system's inner regions.

\subsection{Collisional evolution of an undisturbed, broad planetesimal disk}
\label{sec:collevo}
To produce the observed dust in the outer belt after several hundred Myr of evolution, an ongoing collisional cascade from large, long-lived planetesimals must be in place \citep[e.g.][]{Wyatt2002}.  Detailed collisional calculations applied to Vega's outer belt \citep{Muller2010} indicate that a collisional cascade ignited by a belt of planetesimals confined between 80-120 au could explain most of the observables, including \textit{Spitzer} MIPS radial profiles \citep{Su2005}, the submillimeter morphology from SCUBA images \citep{Holland1998}, as well as the broadband dust spectrum from IR to mm wavelengths. 
While this model is successful at reproducing most observables (pending its application to the ALMA data), it does not explain why Vega's outer belt has the current extent, since it has the \textit{a priori} assumption that the belt is confined to a 80-120 au ring. Additionally, it assumes that planetesimals are born with sizes as large as $\lesssim100$ km at $\sim100$ au, which if changed could affect the amount of observable dust at Vega's age, and therefore the fit to the data.

A model with a radially confined planetesimal distribution makes sense for narrow belts like the one found around Fomalhaut \citep[e.g.][]{Kalas2005, Macgregor2017}, Vega's twin when considering their similar system age and spectral type. Conversely, the remarkable breadth of Vega's outer belt calls for planetesimals being present (and potentially born) at a wide range of radii from the central star. We therefore consider a model where planetesimals successfully form everywhere in the system, and evolve collisionally without external perturbations, having been either pre-stirred at birth during the planetesimal formation process, or stirred very rapidly after formation, in a timescale much shorter than the system age.

As shown in previous work \citep[e.g.][]{Kennedy2010}, even though planetesimals formed everywhere, collisional evolution produces a radially increasing, relatively shallow surface density of planetesimals and mm grains up to radius $r_{\rm c}$, which corresponds to the location where the largest planetesimals (of size $D_{\rm max}$) have collided once within the age of the system. For a radially decreasing initial planetesimal surface density distribution, like that of the MMSN, this $r_{\rm c}$ is also the radial peak of the surface density of planetesimals, which then decreases with the same slope as the MMSN beyond $r_{\rm c}$. Observable grains follow the same radial surface density distribution as the planetesimals out to $\sim r_{\rm c}$, but instead follow a different surface density distribution, flatter than that of the planetesimals, beyond $r_{\rm c}$ \citep{Marino2017a, Schuppler2016, Geiler2017}.  In general, collisional evolution of a broad planetesimal disk predicts a radially increasing surface density of grains with a break, or knee, at $r_{\rm c}$, which is well approximated by a parametric two power-law model such as the one we fitted to the Vega observations. Therefore, we use our double power law fit results to compare with the predictions from collisional evolution, while reminding the reader that a Gaussian or single power law model can fit the data equally well.

A critical observable is the slope $\gamma$ of the inner edge interior to $r_{\rm c}$, where all bodies in the cascade are in collisional equilibrium. This slope is predicted by analytical models \citep[e.g.][]{Wyatt2007a, Lohne2008} as well as more complex numerical simulations \citep[e.g.][]{Schuppler2016, Marino2017a, Geiler2017} to be around $\gamma\sim$2-2.3. This shallow predicted inner slope is inconsistent with the steeper inner slope derived in our modelling ($\gamma=10^{+5}_{-3}$, with a 3$\sigma$ lower limit of 4.4 derived from its posterior probability distribution). Therefore, we conclude that collisional evolution of an extended planetesimal disk is inconsistent with the steep power law gradient of the inner edge derived from the ALMA data.

\subsection{The planetary hypothesis}
\label{sec:planets}

The inconsistency of the belt inner edge shape with models of the collisional evolution for an extended planetesimal disk suggests the presence of a planet currently truncating the inner edge of the planetesimal distribution, unless the planetesimal disk was born truncated within its natal protoplanetary environment (where the latter may or may not be attributable to planet formation).
The current presence of a planet at the inner edge of the belt may also help reconcile the outer belt morphology with the presence of warm ($\sim$170 K) and hot ($\gtrsim$1000 K) dust in the inner regions of the system. We here discuss two scenarios invoking planets to account for Vega's observed dust populations.

\subsubsection{Inward scattering through a chain of planets}
\label{sec:planetchain}A promising scenario that could produce the hot dust (observed within $\sim0.5$ au of the star) is inward scattering of exocomets.  
For this scenario to be successful, inward-scattered exocomets have to reach the inner regions at a sufficiently high rate. This requires that the mechanism of passing material inward is sufficiently efficient, promoting inward scattering rather than ejection \citep[e.g.][]{Wyatt2017}. Maximal efficiency is reached for closely spaced planet chains \citep{Bonsor2012}, with low planet masses in the super-Earth/Neptune size range \citep{Marino2018a}, but not low enough that encounters lead to accretion rather than scattering (e.g. not below $\sim$0.25 M$_{\oplus}$ at 60 au around Vega).
Not only does the mechanism need to be sufficiently efficient, but - for it to be observable around relatively old stars such as Vega - it also needs to be sustained over a non-negligible fraction of the system lifetime. This necessitates replenishment of the population of objects interacting with the outer planet and getting scattered inward. One way to achieve this is, as proposed by \citet{Bonsor2014}, the outermost planet moving into the planetesimal belt as driven by the planetesimals.

\citet{RaymondBonsor2014} simulated inward scattering by a chain of planets around Vega, with planets at 5-30 au migrating outwards into a belt initially extending between 30 and 120 au. For their simulated planet configurations, the outermost, migrating planet in the chain should be 2.5-20 M$_{\oplus}$ - having migrated from $\sim$30 to $\sim$60 au - to produce inward scattering at a sufficiently high rate. This outermost planet would dynamically clear its chaotic zone and produce the inner edge of the belt as observed by ALMA.
Assuming this planet is on a circular orbit and adopting a belt inner edge equal to $r_{\rm c}$ from the double power law model allows us to set a joint constraint on the planet mass and semimajor axis \citep{Wisdom1980}. Further requiring that chaotic zone clearing has taken place within the age of the system \citep{Shannon2016} allows us to break the mass-semimajor axis degeneracy and constrain the planet mass to be $\gtrsim$6 M$_{\oplus}$ and semimajor axis to be $\lesssim$71 au. These are conservative limits, because if the planet is migrating outwards and continuously resupplying the chaotic zone with material, the timescale on which clearing takes place would need to be shorter. Depending on the migration rate, this could make the required planet mass significantly higher and the semimajor axis smaller.

On the other hand, migration is significantly suppressed if the planet is much more massive than the amount of belt material in its encounter zone \citep[$\sim3.5$ Hill radii, e.g.][]{Kirsh2009}, so the planet cannot be too massive if outward migration is to take place and resupply the outer planet with material for inward scattering. Therefore, while this upper mass limit depends on the unconstrained mass of large planetesimals at the belt's current and past inner edge, it is reasonably likely that in this planet chain scenario, as indicated by \citet{RaymondBonsor2014}, the outermost planet is in the super-Earth/Neptune size range.

As well as the outer belt's inner edge, and the hot dust at $<0.5$ au, this inward scattering could also explain the \textit{Spitzer} and \textit{Herschel} detection of warm dust emission in between the hot dust and the outer belt. In this planet chain scenario, warm dust would arise from scattered material on its way into the system's inner regions. The unresolved mid-IR constraint places most of the emission within $\sim$6$\arcsec$, or $\sim$47 au. While this is smaller than the inner edge of the outer belt and so would suggest a gap between the warm and cold dust, we note that low levels of inward-scattered dust may be present out to the outer belt, but have gone undetected in the ALMA data. For example, the simulations of \citet{Marino2018b} for a chain of 30 M$_{\oplus}$ planets around a Sun-like star, with the outermost at 50 au, can produce sufficient inward transport to explain hot dust levels similar to those observed for Vega. These simulations predict $10^{-10}-10^{-8}$ M$_{\oplus}$ au$^{-2}$ in mm-sized dust, assuming that inward-scattered material inherits the same size distribution as the outer belt. This is below our current ALMA 3$\sigma$ upper limits (3.7-5.9$\times10^{-8}$ at 60-14 au), although Vega specific simulations and predictions are needed to draw more robust conclusions.

In summary, the available data appears consistent with a picture where exocomets from the outer belt are being scattered inward by a chain of planets, with the outermost planet potentially migrating outward. In this context, the outermost planet is constrained by the outer belt's inner edge to be $\gtrsim6$ M$_{\oplus}$ and located at $\lesssim71$ au, and in general to likely be in the super-Earth/Neptune mass range to produce sufficient outward migration and inward scattering.
In this picture, inward-scattered material would 1) include dust, to produce an inward scattered disk which may explain Vega's warm excess, and 2) produce hot dust inward of 0.5 au in the assumption that the mass in large exocomets can be efficiently and rapidly converted into dust, for example by copious sublimation near pericenter \citep[e.g.][]{Sezestre2019}. 

\subsubsection{Lone, massive planet}
Another possibility is that the outer belt's inner edge is being carved by a single giant planet. This planet still needs to be more massive than $\sim$6 M$_{\oplus}$ if it is to carve the inner edge of the outer belt within the system age. This would put e.g. a 1 M$_{\rm Jup}$ planet (assumed to be on a circular orbit) at a semimajor axis of $\sim63$ au. Simulations of giant planets interacting with outer belts produce shallower inner edges \citep[e.g.][]{Chiang2009} compared to simulations with lower mass planets as considered in \S\ref{sec:planetchain}, though this slope is also dependent on the planet's or planetesimals' eccentricity \citep{MustillWyatt2012}. This would suggest that constraining the slope of the inner edge at the same time as its location should set a tighter constraint on the allowed region of planet mass - semimajor axis - eccentricity parameter space. However, no explicit predictions have been made in the literature to date on the dependence of the inner edge slope on planet and belt parameters, and what the detailed functional form should be (e.g. Gaussian versus power-law). Therefore, while a steep inner edge as derived from the double power law model would favour lower mass planets on orbits with lower eccentricities, the presence of a giant planet cannot be ruled out.

Upper limits on massive planets from direct imaging suggest that the presence of brown dwarfs around Vega exterior to about 15 au is unlikely, although these limits are formally dependent on the assumed planet evolution models, and on the adopted system age. Reported limits are $\leq$1-3 M$_{\rm Jup}$ within the outer belt (100-200 au), $\leq$5-15 M$_{\rm Jup}$ interior to the outer belt (15-60 au), and $\leq20$ M$_{\rm Jup}$ interior to 15 au \citep{Heinz2008, Janson2015, Meshkat2018}. The presence of a single or multiple giant planets remains therefore possible within the limits imposed by direct imaging and by the belt's inner edge. A giant planet would also produce an outward scattered disk beyond the belt's inner edge, as was proposed to explain the radial profile of the HR8799 belt \citep{Wyatt2017,Geiler2019}. Due to collisional evolution, a scattered disk would produce a relatively flat grain surface density distribution \citep{Wyatt2010}. This is likely consistent with the constraints from the ALMA Vega data, which indicate a shallow outer power-law slope (see Table \ref{tab:beltfit}). 

The presence of a giant planet would still have to be reconciled with the warm and hot dust populations observed closer to the star. In this scenario, the warm dust could originate from an asteroid belt, interior to the giant planet, as originally proposed by \citet{Su2013}. If moderately eccentric ($e_{\rm pl}\sim0.1-0.2$), the giant planet could then produce the hot dust within $\sim0.5$ au by exciting exocomets within the asteroid belt onto eccentric star-grazing orbits, through inner mean motion resonances \citep[e.g.][]{Faramaz2017}. 
When combined with sublimation in the $<0.5$ au region \citep{Marboeuf2016}, this mechanism may resupply hot dust at sufficiently high levels, though detailed simulations are needed to explore this scenario. A key observable of such an eccentric planet between the belts would be an eccentric cavity and thus a potentially detectable offset of the outer belt's geometric center from the star \citep[e.g.][]{Regaly2018}. This could be readily tested with deeper ALMA observations, and with future JWST observations.

Finally, while the gap between the outer and inner asteroid belt would likely be significantly wider than the planet's chaotic zone, no additional giant planets would be needed. This is because a single, eccentric giant planet could clear material far beyond the chaotic zone boundaries through sweeping resonances during the protoplanetary (gas-rich) phase of evolution \citep{Zheng2017}.

In summary, a single giant planet interior to the outer belt may reproduce the belt's inner edge (pending detailed predictions on the dependence of planet mass on the inner edge slope and functional form) and the rather flat surface density distribution observed in the outer belt (expected from a scattered disk of planetesimals). If eccentric, the planet may also supply exocomets producing the hot dust from an asteroid belt reservoir (through outer mean motion resonances), and have produced the wide gap between the asteroid and outer belt through sweeping resonances within the young protoplanetary disk.

\section{Conclusions}\label{sec:conc}
We present new ALMA 1.34~mm observations of the nearby Vega system using the 12m array and ACA to obtain high sensitivity over a wide range of scales, from $\sim1''$ to $30''$ ($\sim8$ to 230~au). These data detect and resolve the outer cold dust emission belt interferometrically for the first time.  We carried out detailed visibility modelling using several parameterizations for the radial surface density of the belt. The key conclusions are:

\begin{enumerate}

\item The face-on millimeter emission belt has a clearly resolved central cavity, and its surface density can be fit by a Gaussian model or by power law models with a steep inner edge (at 60-80 au). The belt is radially very broad, ranging from 60-80~au to at least 150-200~au (the edge of the usable field of view). The images and models show that the belt surface density peaks in the $\sim$75-120~au region. The central star is also strongly detected (signal-to-noise ratio $\sim$200).

\item We place an upper limit on the 1.34~mm flux density of warm dust in the system inner regions discovered by \textit{Spitzer} and \textit{Herschel} \citep{Su2013}. For a narrow belt with radius 14-40~au, the limit is 120-223~$\mu$Jy, while for extended emission interior to 40~au, the limit is $<324$~$\mu$Jy. These $3\sigma$ upper limits are comparable to the predictions from extrapolations of models fitted to the mid-infrared excess emission.

\item We discuss three potential architectures for the Vega system, informed by the new knowledge of the outer belt properties:
(a) collisional evolution of an extended planetesimal disk, which in the absence of planets results in an inner edge slope that is too shallow and inconsistent with the ALMA observations. Unperturbed collisional evolution can only explain the observed morphology if the belt was born truncated;
(b) a chain of closely spaced planets, with an outermost planet of mass $\gtrsim6$ M$_{\oplus}$ at $\lesssim$70~au truncating the inner edge of the outer belt, and with exocomets being efficiently scattered inward to account for both warm and hot dust in the inner regions;
(c) a lone outer giant planet, with mass up to the limit provided by direct imaging ($\sim$5~$M_{\rm Jup}$ at 50-60~au), that truncates the belt's inner edge and ejects planetesimals to produce a scattered disk within the outer belt. If this planet's orbit is eccentric, then it could create a wide gap between the outer planetesimal belt and a putative asteroid belt, and perturb objects in the asteroid belt inward to generate hot dust near the star.

\end{enumerate}

These three proposed scenarios for the Vega system architecture each have characteristic features, and further constraints on the detailed shape of the millimeter emission belt and the morphology of the mid-infrared emission will help to distinguish among them. Upcoming, resolved observations of warm dust with JWST, SOFIA together with deeper, mosaicked ALMA observations are the most likely to provide the most stringent constraints on the presence of planets in this archetypal planetary system. These efforts should be complemented with planet searches with deeper limits, to $\lesssim$1 M$_{\rm Jup}$ beyond 20 au expected to be achievable with JWST, and inwards of that through long-term astrometric monitoring using ALMA long baseline observations. At the same time, we underline the need for detailed predictions on the \textit{shape}, in addition to the location, of the inner edge of planetesimal belts produced by planet-belt interaction. This will be crucial in interpreting current and upcoming high-resolution observations.

\acknowledgments
This paper is dedicated to Wayne Holland, recently passed away (21/05/2019), for his pioneering work in mm astronomy, and in particular the debris disks around Vega and other nearby stars.

LM acknowledges support from the Smithsonian Institution as a Submillimeter Array (SMA) Fellow. AMH is supported by a Cottrell Scholar Award from the Research Corporation for Science Advancement. JPM acknowledges research support by the Ministry of Science and Technology of Taiwan under grants MOST104-2628-M-001-004-MY3 and MOST107-2119-M-001-031-MY3, and Academia Sinica under grant AS-IA-106-M03.
This paper makes use of ALMA data ADS/JAO.ALMA\#2015.1.00182. ALMA is a partnership of ESO (representing its member states), NSF (USA) and NINS (Japan), together with NRC (Canada), NSC and ASIAA (Taiwan), and KASI (Republic of Korea), in cooperation with the Republic of Chile. The Joint ALMA Observatory is operated by ESO, AUI/NRAO and NAOJ. The National Radio Astronomy Observatory is a facility of the National Science Foundation operated under cooperative agreement by Associated Universities, Inc.

\facility{ALMA}.
\software{CASA \citep{McMullin2007},
Matplotlib \citep{Hunter2007}}

\bibliographystyle{aasjournal}
\bibliography{lib}

\newcommand{\noop}[1]{}
\begin{thebibliography}{}
\expandafter\ifx\csname natexlab\endcsname\relax\def\natexlab#1{#1}\fi
\providecommand{\url}[1]{\href{#1}{#1}}

\bibitem[{{Absil} {et~al.}(2006){Absil}, {di Folco}, {M{\'e}rand}, {Augereau},
  {Coud{\'e} du Foresto}, {Aufdenberg}, {Kervella}, {Ridgway}, {Berger}, {ten
  Brummelaar}, {Sturmann}, {Sturmann}, {Turner}, \& {McAlister}}]{Absil2006}
{Absil}, O., {di Folco}, E., {M{\'e}rand}, A., {et~al.} 2006, \aap, 452, 237

\bibitem[{{Allard}(2014)}]{Allard2014}
{Allard}, F. 2014, in IAU Symposium, Vol. 299, Exploring the Formation and
  Evolution of Planetary Systems, ed. M.~{Booth}, B.~C. {Matthews}, \& J.~R.
  {Graham}, 271--272

\bibitem[{{Aufdenberg} {et~al.}(2006){Aufdenberg}, {M{\'e}rand}, {Coud{\'e} du
  Foresto}, {Absil}, {Di Folco}, {Kervella}, {Ridgway}, {Berger}, {ten
  Brummelaar}, {McAlister}, {Sturmann}, {Sturmann}, \&
  {Turner}}]{Aufdenberg2006}
{Aufdenberg}, J.~P., {M{\'e}rand}, A., {Coud{\'e} du Foresto}, V., {et~al.}
  2006, \apj, 645, 664

\bibitem[{{Aumann} {et~al.}(1984){Aumann}, {Gillett}, {Beichman}, {de Jong},
  {Houck}, {Low}, {Neugebauer}, {Walker}, \& {Wesselius}}]{Aumann1984}
{Aumann}, H.~H., {Gillett}, F.~C., {Beichman}, C.~A., {et~al.} 1984, \apjl,
  278, L23

\bibitem[{{Bohlin} \& {Gilliland}(2004)}]{BohlinGilliland2004}
{Bohlin}, R.~C., \& {Gilliland}, R.~L. 2004, \aj, 127, 3508

\bibitem[{{Bonsor} {et~al.}(2012){Bonsor}, {Augereau}, \&
  {Th{\'e}bault}}]{Bonsor2012}
{Bonsor}, A., {Augereau}, J.-C., \& {Th{\'e}bault}, P. 2012, \aap, 548, A104

\bibitem[{{Bonsor} {et~al.}(2014){Bonsor}, {Raymond}, {Augereau}, \&
  {Ormel}}]{Bonsor2014}
{Bonsor}, A., {Raymond}, S.~N., {Augereau}, J.-C., \& {Ormel}, C.~W. 2014,
  \mnras, 441, 2380

\bibitem[{{Booth} {et~al.}(2013){Booth}, {Kennedy}, {Sibthorpe}, {Matthews},
  {Wyatt}, {Duch{\^e}ne}, {Kavelaars}, {Rodriguez}, {Greaves}, {Koning},
  {Vican}, {Rieke}, {Su}, {Moro-Mart{\'{\i}}n}, \& {Kalas}}]{Booth2013}
{Booth}, M., {Kennedy}, G., {Sibthorpe}, B., {et~al.} 2013, \mnras, 428, 1263

\bibitem[{{Carniani} {et~al.}(2015){Carniani}, {Maiolino}, {De Zotti},
  {Negrello}, {Marconi}, {Bothwell}, {Capak}, {Carilli}, {Castellano},
  {Cristiani}, {Ferrara}, {Fontana}, {Gallerani}, {Jones}, {Ohta}, {Ota},
  {Pentericci}, {Santini}, {Sheth}, {Vallini}, {Vanzella}, {Wagg}, \&
  {Williams}}]{Carniani2015}
{Carniani}, S., {Maiolino}, R., {De Zotti}, G., {et~al.} 2015, \aap, 584, A78

\bibitem[{{Chiang} {et~al.}(2009){Chiang}, {Kite}, {Kalas}, {Graham}, \&
  {Clampin}}]{Chiang2009}
{Chiang}, E., {Kite}, E., {Kalas}, P., {Graham}, J.~R., \& {Clampin}, M. 2009,
  \apj, 693, 734

\bibitem[{{Defr{\`e}re} {et~al.}(2011){Defr{\`e}re}, {Absil}, {Augereau}, {di
  Folco}, {Berger}, {Coud{\'e} du Foresto}, {Kervella}, {Le Bouquin},
  {Lebreton}, {Millan-Gabet}, {Monnier}, {Olofsson}, \& {Traub}}]{Defrere2011}
{Defr{\`e}re}, D., {Absil}, O., {Augereau}, J.~C., {et~al.} 2011, \aap, 534, A5

\bibitem[{{Draine}(1978)}]{Draine1978}
{Draine}, B.~T. 1978, \apjs, 36, 595

\bibitem[{{Eiroa} {et~al.}(2013){Eiroa}, {Marshall}, {Mora}, {Montesinos},
  {Absil}, {Augereau}, {Bayo}, {Bryden}, {Danchi}, {del Burgo}, {Ertel},
  {Fridlund}, {Heras}, {Krivov}, {Launhardt}, {Liseau}, {L{\"o}hne},
  {Maldonado}, {Pilbratt}, {Roberge}, {Rodmann}, {Sanz-Forcada}, {Solano},
  {Stapelfeldt}, {Th{\'e}bault}, {Wolf}, {Ardila}, {Ar{\'e}valo}, {Beichmann},
  {Faramaz}, {Gonz{\'a}lez-Garc{\'{\i}}a}, {Guti{\'e}rrez}, {Lebreton},
  {Mart{\'{\i}}nez-Arn{\'a}iz}, {Meeus}, {Montes}, {Olofsson}, {Su}, {White},
  {Barrado}, {Fukagawa}, {Gr{\"u}n}, {Kamp}, {Lorente}, {Morbidelli},
  {M{\"u}ller}, {Mutschke}, {Nakagawa}, {Ribas}, \& {Walker}}]{Eiroa2013}
{Eiroa}, C., {Marshall}, J.~P., {Mora}, A., {et~al.} 2013, \aap, 555, A11

\bibitem[{{Ertel} {et~al.}(2018){Ertel}, {Defr{\`e}re}, {Hinz}, {Mennesson},
  {Kennedy}, {Danchi}, {Gelino}, {Hill}, {Hoffmann}, {Rieke}, {Shannon},
  {Spalding}, {Stone}, {Vaz}, {Weinberger}, {Willems}, {Absil}, {Arbo},
  {Bailey}, {Beichman}, {Bryden}, {Downey}, {Durney}, {Esposito}, {Gaspar},
  {Grenz}, {Haniff}, {Leisenring}, {Marion}, {McMahon}, {Millan-Gabet},
  {Montoya}, {Morzinski}, {Pinna}, {Power}, {Puglisi}, {Roberge}, {Serabyn},
  {Skemer}, {Stapelfeldt}, {Su}, {Vaitheeswaran}, \& {Wyatt}}]{Ertel2018}
{Ertel}, S., {Defr{\`e}re}, D., {Hinz}, P., {et~al.} 2018, \aj, 155, 194

\bibitem[{{Ertel} {et~al.}(2020){Ertel}, {Defr{\`e}re}, {Hinz}, {Mennesson},
  {Kennedy}, {Danchi}, {Gelino}, {Hill}, {Hoffmann}, {Mazoyer}, {Rieke},
  {Shannon}, {Stapelfeldt}, {Spalding}, {Stone}, {Vaz}, {Weinberger},
  {Willems}, {Absil}, {Arbo}, {Bailey}, {Beichman}, {Bryden}, {Downey},
  {Durney}, {Esposito}, {Gaspar}, {Grenz}, {Haniff}, {Leisenring}, {Marion},
  {McMahon}, {Millan-Gabet}, {Montoya}, {Morzinski}, {Perera}, {Pinna}, {Pott},
  {Power}, {Puglisi}, {Roberge}, {Serabyn}, {Skemer}, {Su}, {Vaitheeswaran}, \&
  {Wyatt}}]{Ertel2020}
---. 2020, \aj, 159, 177

\bibitem[{{Faramaz} {et~al.}(2017){Faramaz}, {Ertel}, {Booth}, {Cuadra}, \&
  {Simmonds}}]{Faramaz2017}
{Faramaz}, V., {Ertel}, S., {Booth}, M., {Cuadra}, J., \& {Simmonds}, C. 2017,
  \mnras, 465, 2352

\bibitem[{{Foreman-Mackey} {et~al.}(2013){Foreman-Mackey}, {Hogg}, {Lang}, \&
  {Goodman}}]{Foreman-Mackey2013}
{Foreman-Mackey}, D., {Hogg}, D.~W., {Lang}, D., \& {Goodman}, J. 2013, \pasp,
  125, 306

\bibitem[{{Foreman-Mackey} {et~al.}(2019){Foreman-Mackey}, {Farr}, {Sinha},
  {Archibald}, {Hogg}, {Sanders}, {Zuntz}, {Williams}, {Nelson}, {de
  Val-Borro}, {Erhardt}, {Pashchenko}, \& {Pla}}]{Foreman-Mackey2019}
{Foreman-Mackey}, D., {Farr}, W., {Sinha}, M., {et~al.} 2019, The Journal of
  Open Source Software, 4, 1864

\bibitem[{{Geiler} \& {Krivov}(2017)}]{Geiler2017}
{Geiler}, F., \& {Krivov}, A.~V. 2017, \mnras, 468, 959

\bibitem[{{Geiler} {et~al.}(2019){Geiler}, {Krivov}, {Booth}, \&
  {L{\"o}hne}}]{Geiler2019}
{Geiler}, F., {Krivov}, A.~V., {Booth}, M., \& {L{\"o}hne}, T. 2019, \mnras,
  483, 332

\bibitem[{{Gontcharov}(2006)}]{Gontcharov2006}
{Gontcharov}, G.~A. 2006, Astronomy Letters, 32, 759

\bibitem[{Goodman \& Weare(2010)}]{GoodmanWeare2010}
Goodman, J., \& Weare, J. 2010, Commun. Appl. Math. Comput. Sci., 5, 65.
\newblock \url{http://dx.doi.org/10.2140/camcos.2010.5.65}

\bibitem[{{Gray} \& {Garrison}(1987)}]{GrayGarrison1987}
{Gray}, R.~O., \& {Garrison}, R.~F. 1987, \apjs, 65, 581

\bibitem[{{Heays} {et~al.}(2017){Heays}, {Bosman}, \& {van
  Dishoeck}}]{Heays2017}
{Heays}, A.~N., {Bosman}, A.~D., \& {van Dishoeck}, E.~F. 2017, \aap, 602, A105

\bibitem[{{Heinze} {et~al.}(2008){Heinze}, {Hinz}, {Kenworthy}, {Miller}, \&
  {Sivanandam}}]{Heinz2008}
{Heinze}, A.~N., {Hinz}, P.~M., {Kenworthy}, M., {Miller}, D., \& {Sivanandam},
  S. 2008, \apj, 688, 583

\bibitem[{{Holland} {et~al.}(1998){Holland}, {Greaves}, {Zuckerman}, {Webb},
  {McCarthy}, {Coulson}, {Walther}, {Dent}, {Gear}, \& {Robson}}]{Holland1998}
{Holland}, W.~S., {Greaves}, J.~S., {Zuckerman}, B., {et~al.} 1998, \nat, 392,
  788

\bibitem[{{Holland} {et~al.}(2017){Holland}, {Matthews}, {Kennedy}, {Greaves},
  {Wyatt}, {Booth}, {Bastien}, {Bryden}, {Butner}, {Chen}, {Chrysostomou},
  {Davies}, {Dent}, {Di Francesco}, {Duch{\^e}ne}, {Gibb}, {Friberg}, {Ivison},
  {Jenness}, {Kavelaars}, {Lawler}, {Lestrade}, {Marshall}, {Moro-Martin},
  {Pani{\'c}}, {Phillips}, {Serjeant}, {Schieven}, {Sibthorpe}, {Vican},
  {Ward-Thompson}, {van der Werf}, {White}, {Wilner}, \&
  {Zuckerman}}]{Holland2017}
{Holland}, W.~S., {Matthews}, B.~C., {Kennedy}, G.~M., {et~al.} 2017, \mnras,
  470, 3606

\bibitem[{{Hughes} {et~al.}(2018){Hughes}, {Duch{\^e}ne}, \&
  {Matthews}}]{Hughes2018}
{Hughes}, A.~M., {Duch{\^e}ne}, G., \& {Matthews}, B.~C. 2018, \araa, 56, 541

\bibitem[{{Hughes} {et~al.}(2007){Hughes}, {Wilner}, {Calvet}, {D'Alessio},
  {Claussen}, \& {Hogerheijde}}]{Hughes2007}
{Hughes}, A.~M., {Wilner}, D.~J., {Calvet}, N., {et~al.} 2007, \apj, 664, 536

\bibitem[{{Hughes} {et~al.}(2012){Hughes}, {Wilner}, {Mason}, {Carpenter},
  {Plambeck}, {Chiang}, {Andrews}, {Williams}, {Hales}, {Su}, {Chiang},
  {Dicker}, {Korngut}, \& {Devlin}}]{Hughes2012}
{Hughes}, A.~M., {Wilner}, D.~J., {Mason}, B., {et~al.} 2012, \apj, 750, 82

\bibitem[{Hunter(2007)}]{Hunter2007}
Hunter, J.~D. 2007, Computing in Science \& Engineering, 9, 90

\bibitem[{{Janson} {et~al.}(2015){Janson}, {Quanz}, {Carson}, {Thalmann},
  {Lafreni{\`e}re}, \& {Amara}}]{Janson2015}
{Janson}, M., {Quanz}, S.~P., {Carson}, J.~C., {et~al.} 2015, \aap, 574, A120

\bibitem[{Kalas {et~al.}(2005)Kalas, Graham, \& Clampin}]{Kalas2005}
Kalas, P., Graham, J.~R., \& Clampin, M. 2005, Nature, 435, 1067.
\newblock \url{http://www.ncbi.nlm.nih.gov/pubmed/15973402}

\bibitem[{{Kennedy} {et~al.}(2018){Kennedy}, {Marino}, {Matr{\`a}},
  {Pani{\'c}}, {Wilner}, {Wyatt}, \& {Yelverton}}]{Kennedy2018}
{Kennedy}, G.~M., {Marino}, S., {Matr{\`a}}, L., {et~al.} 2018, \mnras, 475,
  4924

\bibitem[{{Kennedy} \& {Wyatt}(2010)}]{Kennedy2010}
{Kennedy}, G.~M., \& {Wyatt}, M.~C. 2010, \mnras, 405, 1253

\bibitem[{{Kirsh} {et~al.}(2009){Kirsh}, {Duncan}, {Brasser}, \&
  {Levison}}]{Kirsh2009}
{Kirsh}, D.~R., {Duncan}, M., {Brasser}, R., \& {Levison}, H.~F. 2009, \icarus,
  199, 197

\bibitem[{{Koerner} {et~al.}(2001){Koerner}, {Sargent}, \&
  {Ostroff}}]{Koerner2001}
{Koerner}, D.~W., {Sargent}, A.~I., \& {Ostroff}, N.~A. 2001, \apjl, 560, L181

\bibitem[{{L{\"o}hne} {et~al.}(2008){L{\"o}hne}, {Krivov}, \&
  {Rodmann}}]{Lohne2008}
{L{\"o}hne}, T., {Krivov}, A.~V., \& {Rodmann}, J. 2008, \apj, 673, 1123

\bibitem[{{MacGregor} {et~al.}(2015){MacGregor}, {Wilner}, {Andrews}, \&
  {Hughes}}]{Macgregor2015a}
{MacGregor}, M.~A., {Wilner}, D.~J., {Andrews}, S.~M., \& {Hughes}, A.~M. 2015,
  \apj, 801, 59

\bibitem[{{MacGregor} {et~al.}(2017){MacGregor}, {Matr{\`a}}, {Kalas},
  {Wilner}, {Pan}, {Kennedy}, {Wyatt}, {Duchene}, {Hughes}, {Rieke}, {Clampin},
  {Fitzgerald}, {Graham}, {Holland}, {Pani{\'c}}, {Shannon}, \&
  {Su}}]{Macgregor2017}
{MacGregor}, M.~A., {Matr{\`a}}, L., {Kalas}, P., {et~al.} 2017, \apj, 842, 8

\bibitem[{{MacGregor} {et~al.}(2018){MacGregor}, {Weinberger}, {Hughes},
  {Wilner}, {Currie}, {Debes}, {Donaldson}, {Redfield}, {Roberge}, \&
  {Schneider}}]{MacGregor2018}
{MacGregor}, M.~A., {Weinberger}, A.~J., {Hughes}, A.~M., {et~al.} 2018, \apj,
  869, 75

\bibitem[{{Marboeuf} {et~al.}(2016){Marboeuf}, {Bonsor}, \&
  {Augereau}}]{Marboeuf2016}
{Marboeuf}, U., {Bonsor}, A., \& {Augereau}, J.~C. 2016, \planss, 133, 47

\bibitem[{{Marino} {et~al.}(2018{\natexlab{a}}){Marino}, {Bonsor}, {Wyatt}, \&
  {Kral}}]{Marino2018a}
{Marino}, S., {Bonsor}, A., {Wyatt}, M.~C., \& {Kral}, Q. 2018{\natexlab{a}},
  \mnras, 479, 1651

\bibitem[{{Marino} {et~al.}(2019){Marino}, {Yelverton}, {Booth}, {Faramaz},
  {Kennedy}, {Matr{\`a}}, \& {Wyatt}}]{Marino2019}
{Marino}, S., {Yelverton}, B., {Booth}, M., {et~al.} 2019, \mnras, 484, 1257

\bibitem[{{Marino} {et~al.}(2017){Marino}, {Wyatt}, {Pani{\'c}}, {Matr{\`a}},
  {Kennedy}, {Bonsor}, {Kral}, {Dent}, {Duchene}, {Wilner}, {Lisse},
  {Lestrade}, \& {Matthews}}]{Marino2017a}
{Marino}, S., {Wyatt}, M.~C., {Pani{\'c}}, O., {et~al.} 2017, \mnras, 465, 2595

\bibitem[{{Marino} {et~al.}(2018{\natexlab{b}}){Marino}, {Carpenter}, {Wyatt},
  {Booth}, {Casassus}, {Faramaz}, {Guzman}, {Hughes}, {Isella}, {Kennedy},
  {Matr{\`a}}, {Ricci}, \& {Corder}}]{Marino2018b}
{Marino}, S., {Carpenter}, J., {Wyatt}, M.~C., {et~al.} 2018{\natexlab{b}},
  \mnras, 479, 5423

\bibitem[{{Marsh} {et~al.}(2006){Marsh}, {Dowell}, {Velusamy}, {Grogan}, \&
  {Beichman}}]{Marsh2006}
{Marsh}, K.~A., {Dowell}, C.~D., {Velusamy}, T., {Grogan}, K., \& {Beichman},
  C.~A. 2006, \apjl, 646, L77

\bibitem[{Marshall {et~al.}(\noop{3003}in prep.)}]{Marshallinprep}
Marshall, J., {et~al.} \noop{3003}in prep.

\bibitem[{{Matr{\`a}} {et~al.}(2018{\natexlab{a}}){Matr{\`a}}, {Marino},
  {Kennedy}, {Wyatt}, {{\"O}berg}, \& {Wilner}}]{Matra2018b}
{Matr{\`a}}, L., {Marino}, S., {Kennedy}, G.~M., {et~al.} 2018{\natexlab{a}},
  \apj, 859, 72

\bibitem[{{Matr{\`a}} {et~al.}(2015){Matr{\`a}}, {Pani{\'c}}, {Wyatt}, \&
  {Dent}}]{Matra2015}
{Matr{\`a}}, L., {Pani{\'c}}, O., {Wyatt}, M.~C., \& {Dent}, W.~R.~F. 2015,
  \mnras, 447, 3936

\bibitem[{{Matr{\`a}} {et~al.}(2018{\natexlab{b}}){Matr{\`a}}, {Wilner},
  {{\"O}berg}, {Andrews}, {Loomis}, {Wyatt}, \& {Dent}}]{Matra2018a}
{Matr{\`a}}, L., {Wilner}, D.~J., {{\"O}berg}, K.~I., {et~al.}
  2018{\natexlab{b}}, \apj, 853, 147

\bibitem[{{Matr{\`a}} {et~al.}(2019){Matr{\`a}}, {Wyatt}, {Wilner}, {Dent},
  {Marino}, {Kennedy}, \& {Milli}}]{Matra2019b}
{Matr{\`a}}, L., {Wyatt}, M.~C., {Wilner}, D.~J., {et~al.} 2019, \aj, 157, 135

\bibitem[{{Matr{\`a}} {et~al.}(2017){Matr{\`a}}, {MacGregor}, {Kalas}, {Wyatt},
  {Kennedy}, {Wilner}, {Duchene}, {Hughes}, {Pan}, {Shannon}, {Clampin},
  {Fitzgerald}, {Graham}, {Holland}, {Pani{\'c}}, \& {Su}}]{Matra2017b}
{Matr{\`a}}, L., {MacGregor}, M.~A., {Kalas}, P., {et~al.} 2017, \apj, 842, 9

\bibitem[{{McMullin} {et~al.}(2007){McMullin}, {Waters}, {Schiebel}, {Young},
  \& {Golap}}]{McMullin2007}
{McMullin}, J.~P., {Waters}, B., {Schiebel}, D., {Young}, W., \& {Golap}, K.
  2007, in Astronomical Society of the Pacific Conference Series, Vol. 376,
  Astronomical Data Analysis Software and Systems XVI, ed. R.~A. {Shaw},
  F.~{Hill}, \& D.~J. {Bell}, 127

\bibitem[{{Mennesson} {et~al.}(2011){Mennesson}, {Serabyn}, {Hanot}, {Martin},
  {Liewer}, \& {Mawet}}]{Mennesson2011}
{Mennesson}, B., {Serabyn}, E., {Hanot}, C., {et~al.} 2011, \apj, 736, 14

\bibitem[{{Meshkat} {et~al.}(2018){Meshkat}, {Nilsson}, {Aguilar}, {Vasisht},
  {Oppenheimer}, {Su}, {Cady}, {Lockhart}, {Matthews}, {Dekany}, {Leisenring},
  {Ygouf}, {Mawet}, {Pueyo}, \& {Beichman}}]{Meshkat2018}
{Meshkat}, T., {Nilsson}, R., {Aguilar}, J., {et~al.} 2018, \aj, 156, 214

\bibitem[{{Monnier} {et~al.}(2012){Monnier}, {Che}, {Zhao}, {Ekstr{\"o}m},
  {Maestro}, {Aufdenberg}, {Baron}, {Georgy}, {Kraus}, {McAlister}, {Pedretti},
  {Ridgway}, {Sturmann}, {Sturmann}, {ten Brummelaar}, {Thureau}, {Turner}, \&
  {Tuthill}}]{Monnier2012}
{Monnier}, J.~D., {Che}, X., {Zhao}, M., {et~al.} 2012, \apjl, 761, L3

\bibitem[{{Morales} {et~al.}(2016){Morales}, {Bryden}, {Werner}, \&
  {Stapelfeldt}}]{Morales2016}
{Morales}, F.~Y., {Bryden}, G., {Werner}, M.~W., \& {Stapelfeldt}, K.~R. 2016,
  \apj, 831, 97

\bibitem[{{M{\"u}ller} {et~al.}(2010){M{\"u}ller}, {L{\"o}hne}, \&
  {Krivov}}]{Muller2010}
{M{\"u}ller}, S., {L{\"o}hne}, T., \& {Krivov}, A.~V. 2010, \apj, 708, 1728

\bibitem[{{Mustill} \& {Wyatt}(2012)}]{MustillWyatt2012}
{Mustill}, A.~J., \& {Wyatt}, M.~C. 2012, \mnras, 419, 3074

\bibitem[{{Pawellek} \& {Krivov}(2015)}]{Pawellek2015}
{Pawellek}, N., \& {Krivov}, A.~V. 2015, \mnras, 454, 3207

\bibitem[{{Pi{\'e}tu} {et~al.}(2011){Pi{\'e}tu}, {di Folco}, {Guilloteau},
  {Gueth}, \& {Cox}}]{Pietu2011}
{Pi{\'e}tu}, V., {di Folco}, E., {Guilloteau}, S., {Gueth}, F., \& {Cox}, P.
  2011, \aap, 531, L2

\bibitem[{{Raymond} \& {Bonsor}(2014)}]{RaymondBonsor2014}
{Raymond}, S.~N., \& {Bonsor}, A. 2014, \mnras, 442, L18

\bibitem[{{Reg{\'a}ly} {et~al.}(2018){Reg{\'a}ly}, {Dencs}, {Mo{\'o}r}, \&
  {Kov{\'a}cs}}]{Regaly2018}
{Reg{\'a}ly}, Z., {Dencs}, Z., {Mo{\'o}r}, A., \& {Kov{\'a}cs}, T. 2018,
  \mnras, 473, 3547

\bibitem[{{Schneider} {et~al.}(2014){Schneider}, {Grady}, {Hines}, {Stark},
  {Debes}, {Carson}, {Kuchner}, {Perrin}, {Weinberger}, {Wisniewski},
  {Silverstone}, {Jang-Condell}, {Henning}, {Woodgate}, {Serabyn},
  {Moro-Martin}, {Tamura}, {Hinz}, \& {Rodigas}}]{Schneider2014}
{Schneider}, G., {Grady}, C.~A., {Hines}, D.~C., {et~al.} 2014, \aj, 148, 59

\bibitem[{{Sch{\"u}ppler} {et~al.}(2016){Sch{\"u}ppler}, {Krivov}, {L{\"o}hne},
  {Booth}, {Kirchschlager}, \& {Wolf}}]{Schuppler2016}
{Sch{\"u}ppler}, C., {Krivov}, A.~V., {L{\"o}hne}, T., {et~al.} 2016, \mnras,
  461, 2146

\bibitem[{{Sezestre} {et~al.}(2019){Sezestre}, {Augereau}, \&
  {Th{\'e}bault}}]{Sezestre2019}
{Sezestre}, {\'E}., {Augereau}, J.~C., \& {Th{\'e}bault}, P. 2019, \aap, 626,
  A2

\bibitem[{{Shannon} {et~al.}(2016){Shannon}, {Wu}, \& {Lithwick}}]{Shannon2016}
{Shannon}, A., {Wu}, Y., \& {Lithwick}, Y. 2016, \apj, 818, 175

\bibitem[{{Sibthorpe} {et~al.}(2010){Sibthorpe}, {Vandenbussche}, {Greaves},
  {Pantin}, {Olofsson}, {Acke}, {Barlow}, {Blommaert}, {Bouwman}, {Brandeker},
  {Cohen}, {De Meester}, {Dent}, {di Francesco}, {Dominik}, {Fridlund}, {Gear},
  {Glauser}, {Gomez}, {Hargrave}, {Harvey}, {Henning}, {Heras}, {Hogerheijde},
  {Holland}, {Ivison}, {Leeks}, {Lim}, {Liseau}, {Matthews}, {Naylor},
  {Pilbratt}, {Polehampton}, {Regibo}, {Royer}, {Sicilia-Aguilar}, {Swinyard},
  {Waelkens}, {Walker}, \& {Wesson}}]{Sibthorpe2010}
{Sibthorpe}, B., {Vandenbussche}, B., {Greaves}, J.~S., {et~al.} 2010, \aap,
  518, L130

\bibitem[{{Su} {et~al.}(2005){Su}, {Rieke}, {Misselt}, {Stansberry},
  {Moro-Martin}, {Stapelfeldt}, {Werner}, {Trilling}, {Bendo}, {Gordon},
  {Hines}, {Wyatt}, {Holland }, {Marengo}, {Megeath}, \& {Fazio}}]{Su2005}
{Su}, K.~Y.~L., {Rieke}, G.~H., {Misselt}, K.~A., {et~al.} 2005, \apj, 628, 487

\bibitem[{{Su} {et~al.}(2013){Su}, {Rieke}, {Malhotra}, {Stapelfeldt},
  {Hughes}, {Bonsor}, {Wilner}, {Balog}, {Watson}, {Werner}, \&
  {Misselt}}]{Su2013}
{Su}, K.~Y.~L., {Rieke}, G.~H., {Malhotra}, R., {et~al.} 2013, \apj, 763, 118

\bibitem[{{Su} {et~al.}(2017){Su}, {MacGregor}, {Booth}, {Wilner}, {Flaherty},
  {Hughes}, {Phillips}, {Malhotra}, {Hales}, {Morrison}, {Ertel}, {Matthews},
  {Dent}, \& {Casassus}}]{Su2017}
{Su}, K. Y.~L., {MacGregor}, M.~A., {Booth}, M., {et~al.} 2017, \aj, 154, 225

\bibitem[{{Tazzari} {et~al.}(2018){Tazzari}, {Beaujean}, \&
  {Testi}}]{Tazzari2018}
{Tazzari}, M., {Beaujean}, F., \& {Testi}, L. 2018, \mnras, 476, 4527

\bibitem[{{Thureau} {et~al.}(2014){Thureau}, {Greaves}, {Matthews}, {Kennedy},
  {Phillips}, {Booth}, {Duch{\^e}ne}, {Horner}, {Rodriguez}, {Sibthorpe}, \&
  {Wyatt}}]{thureau2014}
{Thureau}, N.~D., {Greaves}, J.~S., {Matthews}, B.~C., {et~al.} 2014, \mnras,
  445, 2558

\bibitem[{{van Dishoeck} {et~al.}(2006){van Dishoeck}, {Jonkheid}, \& {van
  Hemert}}]{vanDishoeck2006}
{van Dishoeck}, E.~F., {Jonkheid}, B., \& {van Hemert}, M.~C. 2006, Faraday
  Discussions, 133, 231

\bibitem[{{van Leeuwen}(2007)}]{vanLeeuwen2007}
{van Leeuwen}, F. 2007, \aap, 474, 653

\bibitem[{{Visser} {et~al.}(2009){Visser}, {van Dishoeck}, \&
  {Black}}]{Visser2009}
{Visser}, R., {van Dishoeck}, E.~F., \& {Black}, J.~H. 2009, \aap, 503, 323

\bibitem[{{White} {et~al.}(2020){White}, {Tapia-V{\'a}zquez}, {Hughes},
  {Mo{\'o}r}, {Matthews}, {Wilner}, {Aufdenberg}, {Hughes}, {De la Luz}, \&
  {Boley}}]{White2020}
{White}, J.~A., {Tapia-V{\'a}zquez}, F., {Hughes}, A.~G., {et~al.} 2020, \apj,
  894, 76

\bibitem[{{Wilner} {et~al.}(2002){Wilner}, {Holman}, {Kuchner}, \&
  {Ho}}]{Wilner2002}
{Wilner}, D.~J., {Holman}, M.~J., {Kuchner}, M.~J., \& {Ho}, P.~T.~P. 2002,
  \apjl, 569, L115

\bibitem[{{Wilner} {et~al.}(2018){Wilner}, {MacGregor}, {Andrews}, {Hughes},
  {Matthews}, \& {Su}}]{Wilner2018}
{Wilner}, D.~J., {MacGregor}, M.~A., {Andrews}, S.~M., {et~al.} 2018, \apj,
  855, 56

\bibitem[{{Wisdom}(1980)}]{Wisdom1980}
{Wisdom}, J. 1980, \aj, 85, 1122

\bibitem[{{Wyatt}(2006)}]{Wyatt2006}
{Wyatt}, M.~C. 2006, \apj, 639, 1153

\bibitem[{{Wyatt} {et~al.}(2017){Wyatt}, {Bonsor}, {Jackson}, {Marino}, \&
  {Shannon}}]{Wyatt2017}
{Wyatt}, M.~C., {Bonsor}, A., {Jackson}, A.~P., {Marino}, S., \& {Shannon}, A.
  2017, \mnras, 464, 3385

\bibitem[{{Wyatt} {et~al.}(2010){Wyatt}, {Booth}, {Payne}, \&
  {Churcher}}]{Wyatt2010}
{Wyatt}, M.~C., {Booth}, M., {Payne}, M.~J., \& {Churcher}, L.~J. 2010, \mnras,
  402, 657

\bibitem[{{Wyatt} \& {Dent}(2002)}]{Wyatt2002}
{Wyatt}, M.~C., \& {Dent}, W.~R.~F. 2002, \mnras, 334, 589

\bibitem[{{Wyatt} {et~al.}(2007){Wyatt}, {Smith}, {Greaves}, {Beichman},
  {Bryden}, \& {Lisse}}]{Wyatt2007a}
{Wyatt}, M.~C., {Smith}, R., {Greaves}, J.~S., {et~al.} 2007, \apj, 658, 569

\bibitem[{{Yoon} {et~al.}(2010){Yoon}, {Peterson}, {Kurucz}, \&
  {Zagarello}}]{Yoon2010}
{Yoon}, J., {Peterson}, D.~M., {Kurucz}, R.~L., \& {Zagarello}, R.~J. 2010,
  \apj, 708, 71

\bibitem[{{Zheng} {et~al.}(2017){Zheng}, {Lin}, {Kouwenhoven}, {Mao}, \&
  {Zhang}}]{Zheng2017}
{Zheng}, X., {Lin}, D. N.~C., {Kouwenhoven}, M.~B.~N., {Mao}, S., \& {Zhang},
  X. 2017, \apj, 849, 98

\end{thebibliography}

\end{document}